\documentclass{pasj00}
\usepackage{comment}
\draft
\begin{document}
\SetRunningHead{Author(s) in page-head}{Running Head}
\Received{29-Sep-2014}{}
\Accepted{26-Dec-2014}{}

\title{High Dispersion Spectroscopy of Solar-type Superflare Stars. \\ 
II. Stellar Rotation, Starspots, and Chromospheric Activities}

\author{Yuta \textsc{Notsu}\altaffilmark{1}, Satoshi \textsc{Honda}\altaffilmark{2}, Hiroyuki \textsc{Maehara}\altaffilmark{3,4}, 
Shota \textsc{Notsu}\altaffilmark{1},  Takuya \textsc{Shibayama}\altaffilmark{5}, 
Daisaku \textsc{Nogami}\altaffilmark{1,6}, and Kazunari \textsc{Shibata}\altaffilmark{6}} 
    
\email{ynotsu@kwasan.kyoto-u.ac.jp}    
\affil{\altaffilmark{1}Department of Astronomy, Kyoto University, Kitashirakawa-Oiwake-cho,
Sakyo-ku, Kyoto 606-8502}
\affil{\altaffilmark{2}Center for Astronomy, University of Hyogo, 407-2, Nishigaichi, Sayo-cho, Sayo, Hyogo 679-5313}
\affil{\altaffilmark{3}Kiso Observatory, Institute of Astronomy, School of Science, The University of Tokyo, 10762-30, Mitake, 
Kiso-machi, Kiso-gun, Nagano 397-0101}
\affil{\altaffilmark{4}Okayama Astrophysical Observatory, National Astronomical Observatory of Japan, 
3037-5 Honjo, Kamogata, Asakuchi, Okayama 719-0232}
\affil{\altaffilmark{5}Solar-Terrestrial Environment Laboratory, Nagoya University, Furo-cho, Chikusa-ku, Nagoya, Aichi, 464-8601}
\affil{\altaffilmark{6}Kwasan and Hida Observatories, Kyoto University, Yamashina-ku, Kyoto 607-8471}


\maketitle

\begin{abstract}
We conducted high dispersion spectroscopic observations of 50 superflare stars with Subaru/HDS.
These 50 stars were selected from the solar-type superflare stars that we had discovered from the Kepler data.
More than half (34 stars) of these 50 target superflare stars show no evidence of binarity, 
and we estimated stellar parameters of these 34 stars in our previous study (\cite{PaperI}, hereafter called Paper I).  
According to our previous studies using Kepler data, superflare stars show 
quasi-periodic brightness variations whose amplitude (0.1-10\%) is 
much larger than that of the solar brightness variations (0.01-0.1\%) 
caused by the existence of sunspots on the rotating solar surface.
In this study, we investigated whether these quasi-periodic brightness variations of superflare stars 
are explained by the rotation of a star with fairly large starspots,
by using stellar parameters derived in Paper I.
First, we confirmed that the value of the projected rotational velocity $v \sin i$ 
is consistent with the rotational velocity estimated from the period of the brightness variation.
Next, we measured the intensity of Ca II infrared triplet lines and H$\alpha$ line, good indicators of the stellar chromospheric activity, 
and compared them with other stellar properties. 
The intensity of Ca II infrared triplet lines indicates that the mean magnetic field strength ($\langle fB\rangle$) 
of the target superflare stars can be higher than that of the Sun.
A correlation between the amplitude of the brightness variation and 
the intensity of Ca II triplet line was found. 
All the targets expected to have large starspots because of their large amplitude of the brightness variation show high 
chromospheric activities compared to the Sun.
These results support that the brightness variation of superflare stars is due to the rotation with large starspots.  
\end{abstract}

\bigskip
\bigskip
\bigskip
\bigskip

\section{Introduction}\label{sec:intro}
Flares are the energetic explosions in the stellar atmosphere 
and are thought to occur by intense releases of 
magnetic energy stored around starspots, like solar flares (e.g., \cite{Shibata2011}).
The total energy released in the largest solar flares is estimated to be of the order of $10^{32}$ erg (e.g., \cite{Priest1981}; \cite{Emslie2012}).
Many ``superflare" events have been recently found by using the Kepler data (\cite{Maehara2012}; \cite{Shibayama2013}).
Superflares are gigantic flare events that are 10-$10^6$ times more energetic ($\sim10^{33-38}$ erg; \cite{Schaefer2000}) 
than the largest solar flares on the Sun ($\sim10^{32}$erg).
We have analyzed the data from Kepler space telescope \citep{Koch2010},
and discovered 365 superflare events on 148 solar-type (G-type main-sequence) stars from the data of 83,000 solar-type stars 
observed for the first 120 days of the Kepler mission \citep{Maehara2012}.
We here define solar-type stars as the stars that have a surface temperature of $5100\leq T_{\rm{eff}}\leq 6000$K and a surface gravity of $\log g \geq 4.0$.
Extending the study of \citet{Maehara2012}, we found 1547 superflare events on 279 solar-type stars 
by using the Kepler data of a longer period ($\sim$500 days) \citep{Shibayama2013}. 
Kepler is very useful for detecting small increases in the stellar brightness caused by stellar flares, 
because Kepler realized high photometric precision exceeding 0.01\% for moderately bright stars, 
and obtained continuous time-series data of many stars over a long period \citep{Koch2010}.
\\ \\
\ \ \ \ \ \ \
The analyses of Kepler data enabled us to discuss statistical properties of superflares 
since a very large number of flare events were discovered.
\citet{Shibayama2013} confirmed that the occurrence rate ($dN/dE$) of the superflare versus the flare energy ($E$) has a power-law distribution
of $dN/dE\propto E^{-\alpha}$, where $\alpha\sim 2$, and this distribution is roughly similar to that for the solar flare.
\citet{Shibayama2013} also estimated that a superflare with an energy of $10^{34-35}$ erg occurs once in 800-5000 years in Sun-like stars. 
``Sun-like stars" are here defined as solar-type stars with an effective temperature of $5600\leq T_{\rm{eff}}\leq 6000$K, a surface gravity of $\log g \geq 4.0$, 
and a rotation period ($P$) longer than 10 days.
\\ \\
\ \ \ \ \ \ \
Considering these results, we have performed more analyses in order to investigate 
whether properties of superflare stars can be explained by applying our current physical understanding of the Sun.
Solar flares are the intense releases of magnetic energy stored around sunspots, as mentioned above.
If we assume superflares are the similar magnetic energy releases, 
large starspots are necessary to explain their large amount of energy.
Many of the superflare stars show quasi-periodic brightness variations with a typical period of from one day to a few tens of days.
The amplitude of these brightness variations is in the range of 0.1-10\% \citep{Maehara2012}, 
and is much larger than that of the solar brightness variation (0.01-0.1\%; e.g., \cite{Lanza2003}) caused by the existence of sunspots on the rotating solar surface.
Many ordinary solar-type stars observed by Kepler also show such brightness variations 
and these variations are thought to be caused by the rotation of a star having starspots (e.g., \cite{Basri2011}; \cite{Reinhold2013}).
\citet{YNotsu2013} showed that the above brightness variations of superflare stars can be well explained 
by the rotation of a star with fairly large starspots, taking into account the effects of inclination angle and the spot latitude.
In other words, the existence of large starspots is also supported by the brightness variations. 
\\ \\
\ \ \ \ \ \ \
\citet{YNotsu2013} compared the superflare energy and frequency with the rotation period, 
assuming that the brightness variation period corresponds to the rotation period. 
They then found slowly rotating stars can still produce as energetic flares as those of more rapidly rotating stars although the average flare
frequency is lower for more slowly rotating stars.
\citet{YNotsu2013} also clarified that the superflare energy is related to the total coverage of the starspots, 
and that the energy of superflares can be explained by the magnetic energy stored around these large starspots.
In addition, \citet{Shibata2013} suggested, on the basis of theoretical estimates, that the Sun can generate large magnetic flux sufficient for causing
superflares with an energy of $10^{34}$ erg within one solar cycle ($\sim$11yr).
\\ \\
\ \ \ \ \ \ \
The results described above are, however, only based on Kepler monochromatic photometric data. 
We need to spectroscopically investigate whether these brightness variations are explained by the rotation, and whether superflare stars have large starspots.
The stellar parameters and the binarity of the superflare stars are also needed to be investigated with spectroscopic observations
in order to discuss whether the Sun can really generate superflares.
\\ \\
\ \ \ \ \ \ \
We have then performed high dispersion spectroscopy of 50 superflares stars.
We have already reported some results of our spectroscopic observations of three superflare stars 
in \citet{SNotsu2013} and \citet{Nogami2014}.
In particular, \citet{Nogami2014} found that spectroscopic properties ($T_{\rm{eff}}$, $\log g$, [Fe/H], and rotational velocity)
of the two superflare stars KIC9766237 and KIC9944137 are very close to those of the Sun.
In \authorcite{PaperI} (\yearcite{PaperI}; hereinafter referred to as Paper I),
we have already conducted measurements of stellar parameters for the sake of detailed discussions in this paper.
In paper I, we investigated the binarity of the target stars, and found that more than half (34 stars) of 50 target superflare stars have no evidence of binarity. 
We then estimated temperature ($T_{\rm{eff}}$), surface gravity ($\log g$), metallicity ([Fe/H]), and projected rotational velocity ($v\sin i$) 
of these 34 ``single" superflare stars. 
The accuracy of our stellar atmospheric parameters ($T_{\rm{eff}}$, $\log g$, and [Fe/H]) is higher than that of KIC (Kepler Input Catalog; \cite{Brown2011}) values, 
and the differences between our values and KIC values 
($(\Delta T_{\rm{eff}})_{\rm{rms}} \sim 219$K, $(\Delta \log g)_{\rm{rms}} \sim 0.37$ dex, and $(\Delta\rm{[Fe/H]})_{\rm{rms}} \sim 0.46$ dex) 
are comparable to the large uncertainties and systematic differences 
in KIC values reported by the previous researches (e.g., \cite{Brown2011}). 
We confirmed that the estimated temperature and $\log g$ values of these 34 superflare stars are roughly in the range of solar-type (G-type main sequence stars) stars.
In particular, the temperature, surface gravity, and the brightness variation period ($P_{0}$) 
of 9 stars including two stars (KIC9766237 and KIC9944137) reported in \citet{Nogami2014} 
are in the range of ``Sun-like" stars ($5600\leq T_{\rm{eff}}\leq 6000$K, $\log g\geq$4.0, and $P_{0}>$10 days).
We also found that five of the 34 target stars are fast rotators ($v \sin i \geq 10$km s$^{-1}$), 
while 22 stars have relatively low $v \sin i$ values ($v \sin i<5$km s$^{-1}$). 
These results suggest that stars whose spectroscopic properties similar to the Sun can have superflares, 
and this supports the hypothesis that the Sun might cause a superflare.
\\ \\
\ \ \ \ \ \ \
On the basis of the above stellar parameters derived in Paper I, 
in this paper we investigate whether the brightness variation is explained by the rotation, and whether superflare stars have large starspots.
In this process, we use ``$v \sin i$" estimated in Paper I.
Moreover, we use the intensity of Ca II infrared triplet (Ca II IRT, $\lambda=8498,~8542,~8662$\AA) lines and H$\alpha$ line, 
which we have already used for investigating chromospheric activity in \citet{SNotsu2013} and \citet{Nogami2014}. 
Ca II IRT and H$\alpha$ lines are well-known good indicators of the chromospheric activity of G-type stars (e.g., \cite{Takeda2010}; \cite{Soderblom1993b}; \cite{SNotsu2013}).
\\
\\
\ \ \ \ \ \ \
In Section \ref{sec:obsdata}, we briefly explain the observational data used in this paper.
In Section \ref{subsec:CaIIHa}, we show the measurement results of the intensity of Ca II IRT and H$\alpha$ lines.
We then indirectly estimate the mean intensity of stellar magnetic field using Ca II 8542 line on the basis of our solar spectroheliographic observation in Section \ref{subsec:Hida}.
In Section \ref{subsec:inc}, we compare ``$v\sin i$" with the period of the brightness variation, 
and consider whether the brightness variation is explained by the rotation.
In Section \ref{subsec:activity-spots}, we discuss stellar chromospheric activities, including the starspot sizes and rotational velocity of superflare stars. 
\\
\\

\section{Observational Data}\label{sec:obsdata}
\begin{figure}[htbp]
 \begin{center}
  \FigureFile(70mm,70mm){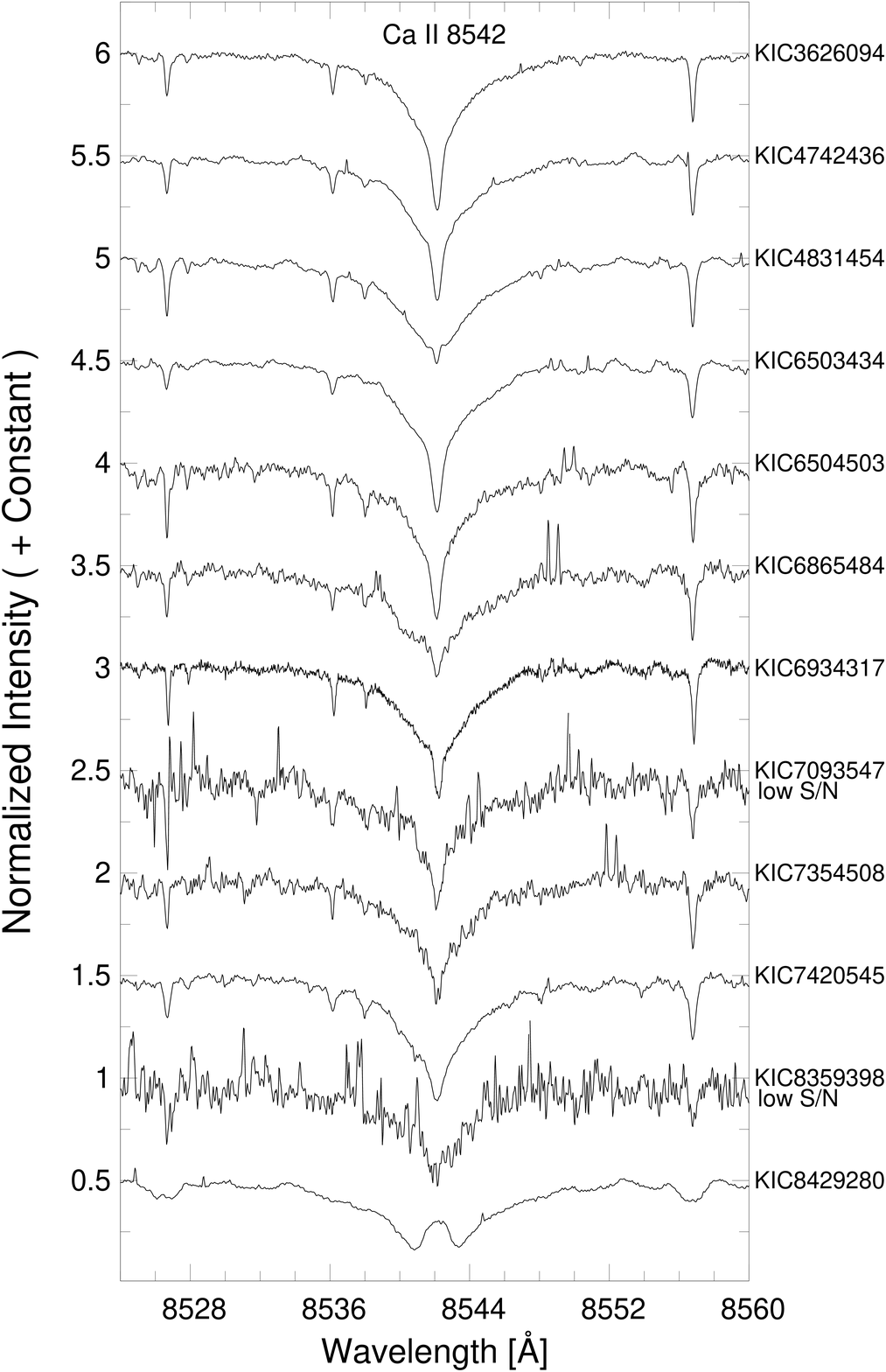}
  \FigureFile(70mm,70mm){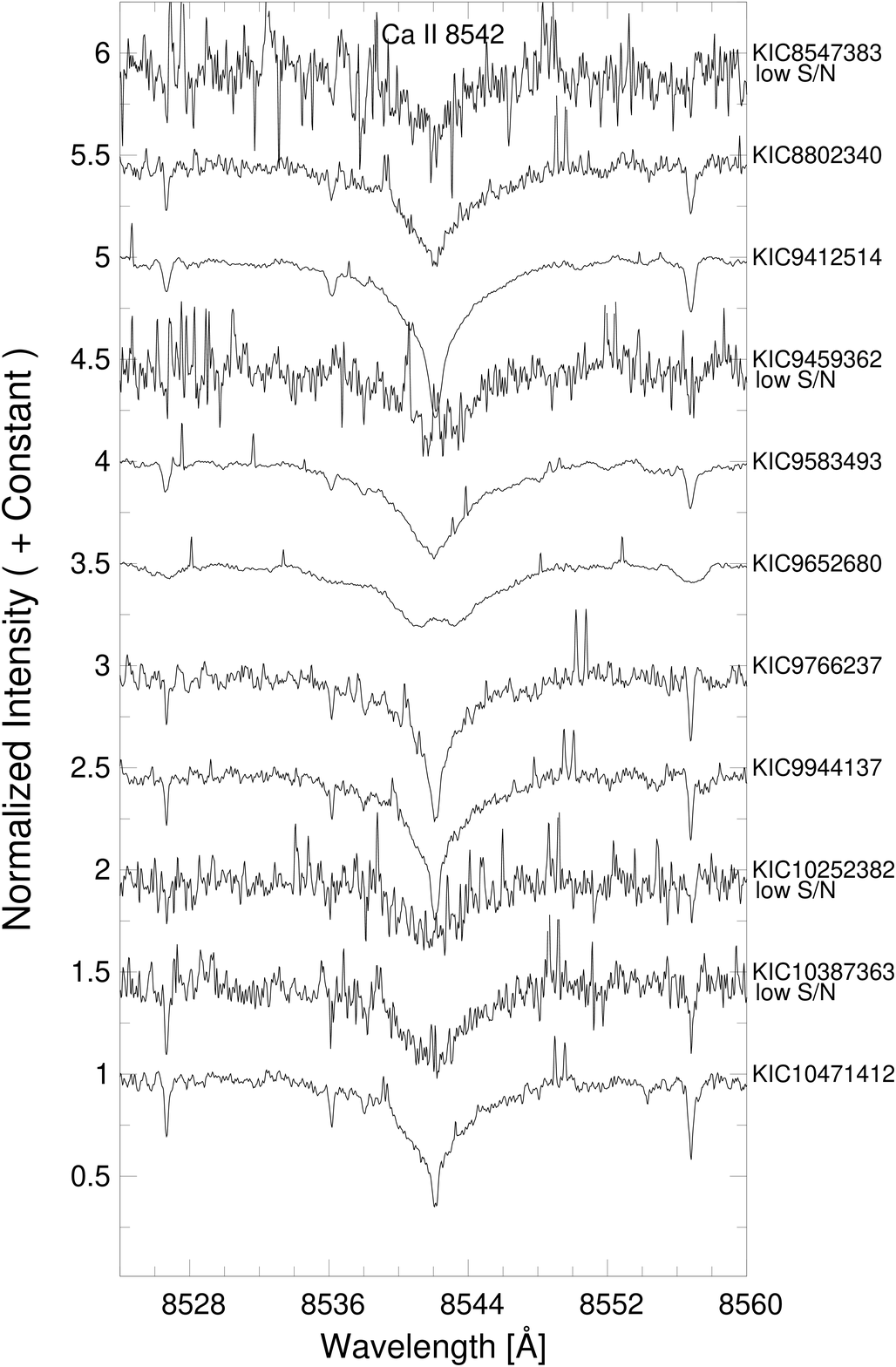}
  \FigureFile(70mm,70mm){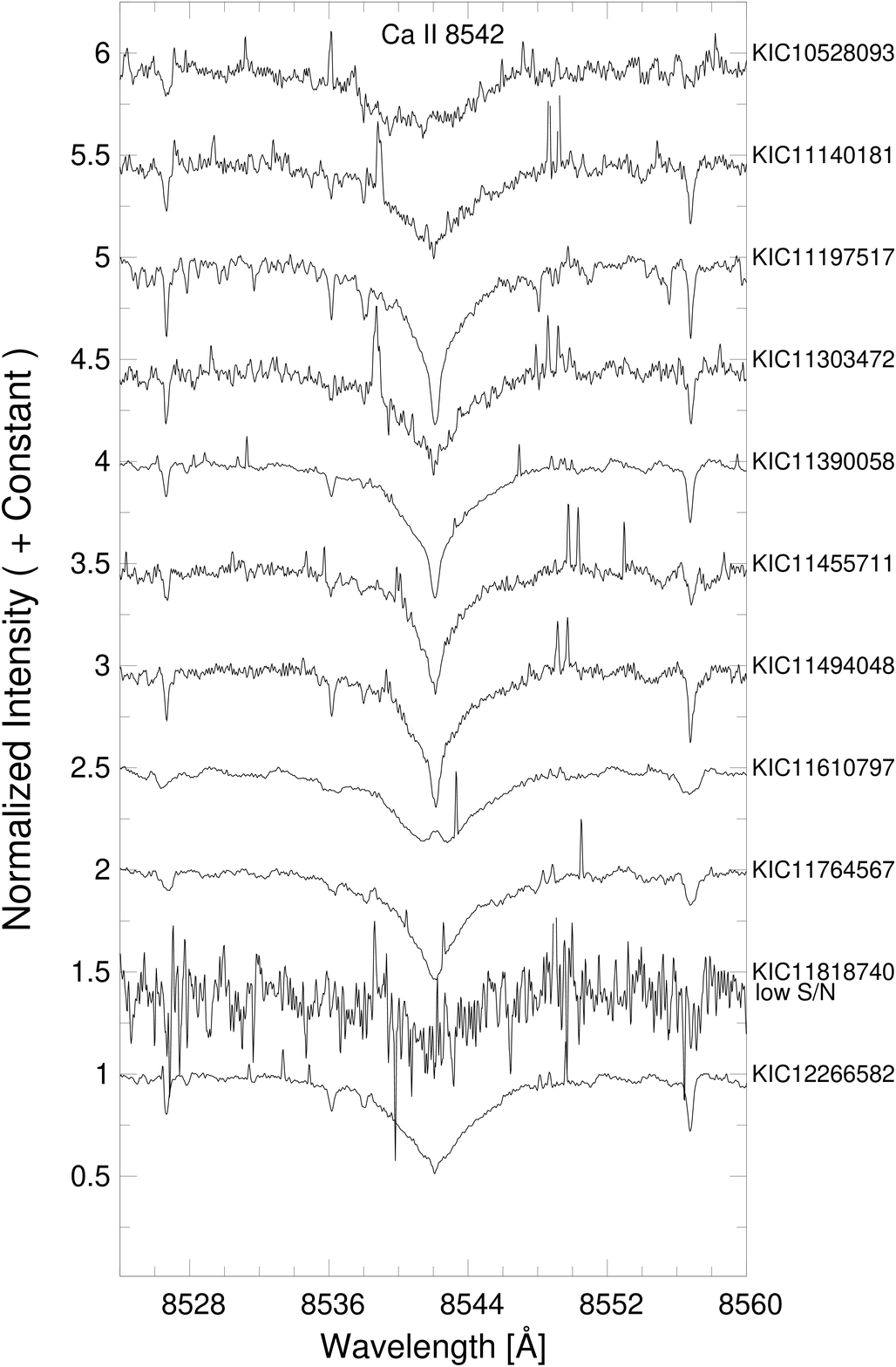}
  \FigureFile(70mm,70mm){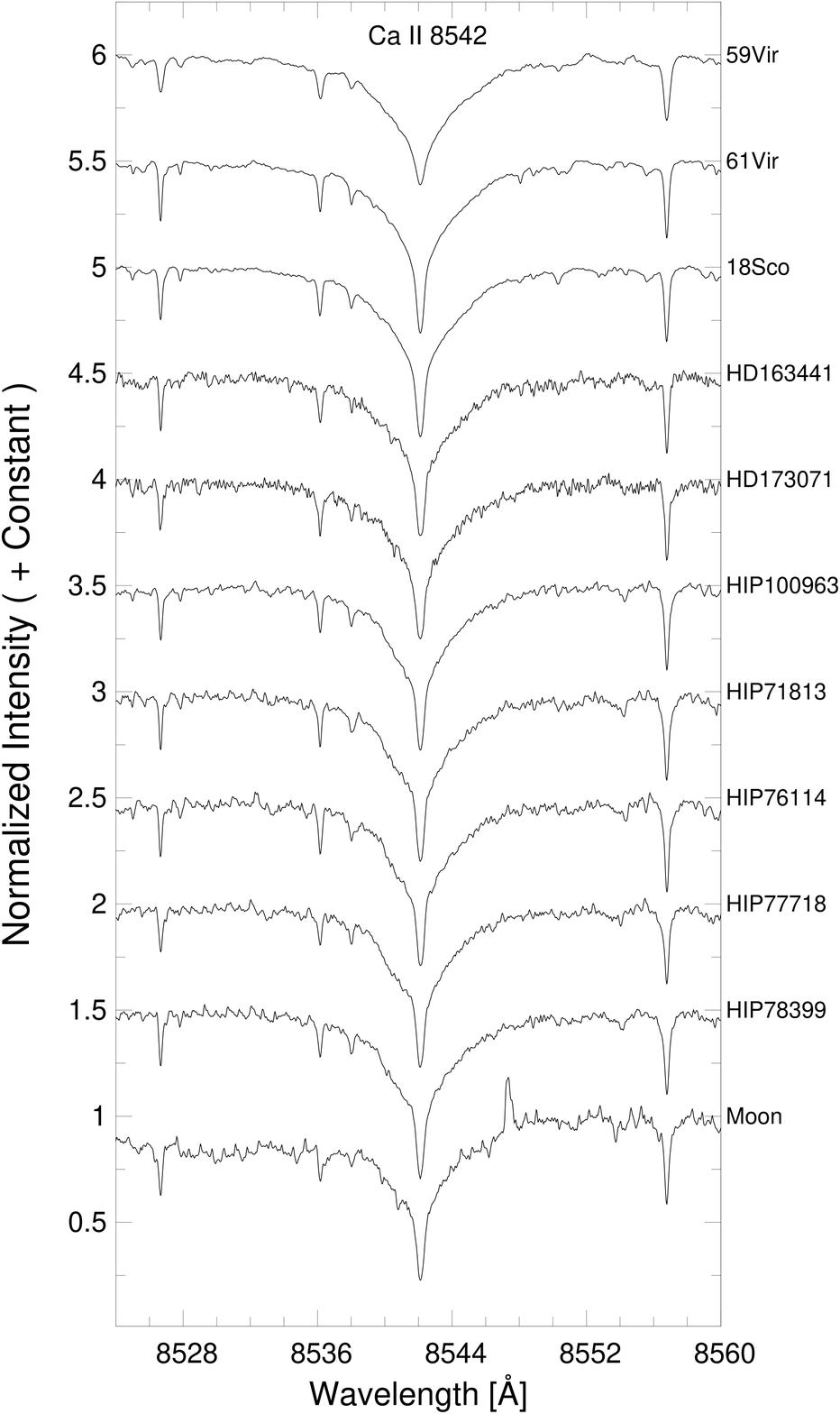}
 \end{center}
\caption{Spectra around Ca II 8542 line of the 34 superflare stars that show no evidence of binarity, 10 comparison stars, and Moon.
The wavelength scale is adjusted to the laboratory frame. Co-added spectra are used here in case that the star was observed multiple times.
}\label{fig:sg1Ca}
\end{figure}

\begin{figure}[htbp]
 \begin{center}
  \FigureFile(70mm,70mm){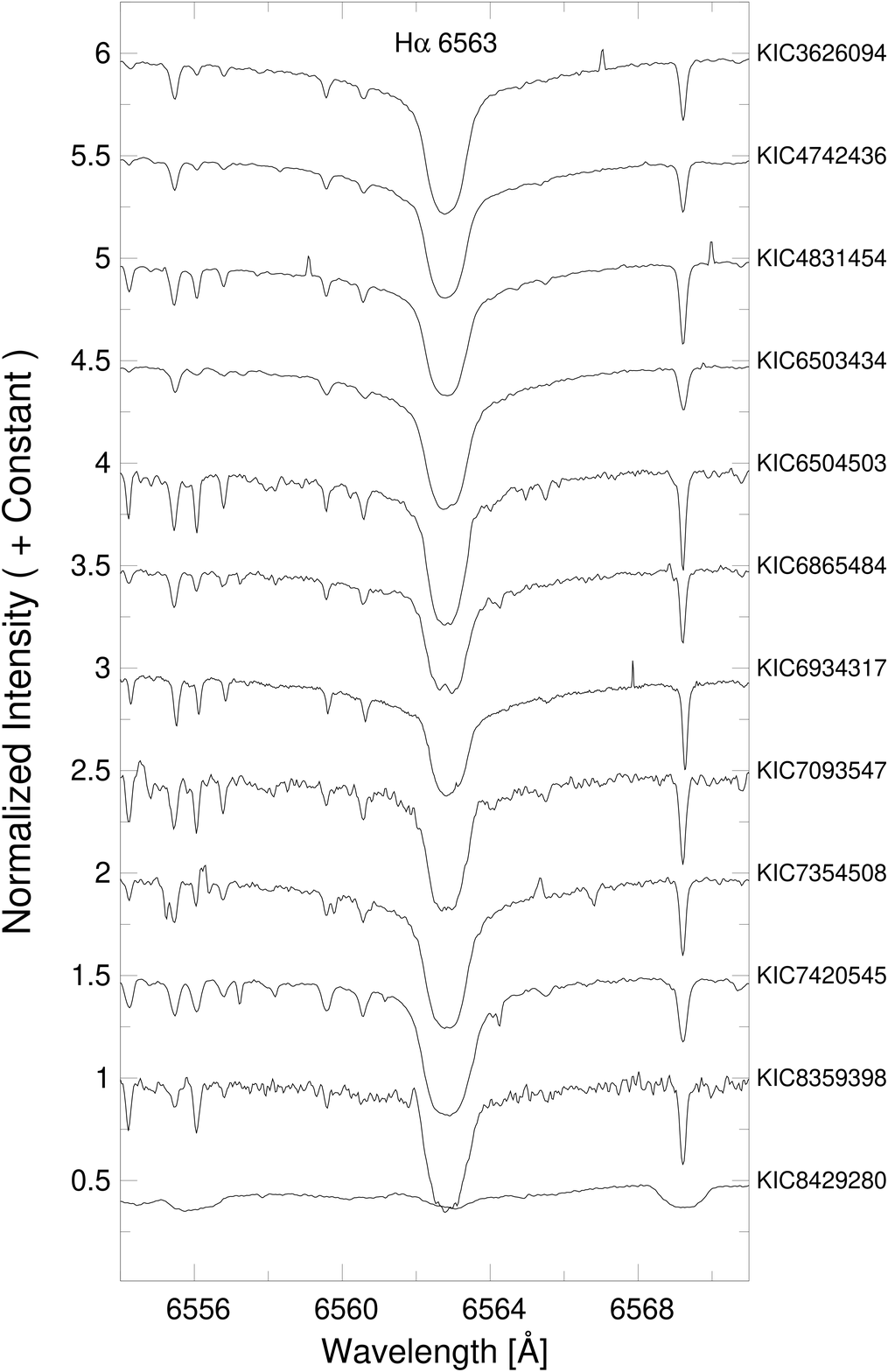}
  \FigureFile(70mm,70mm){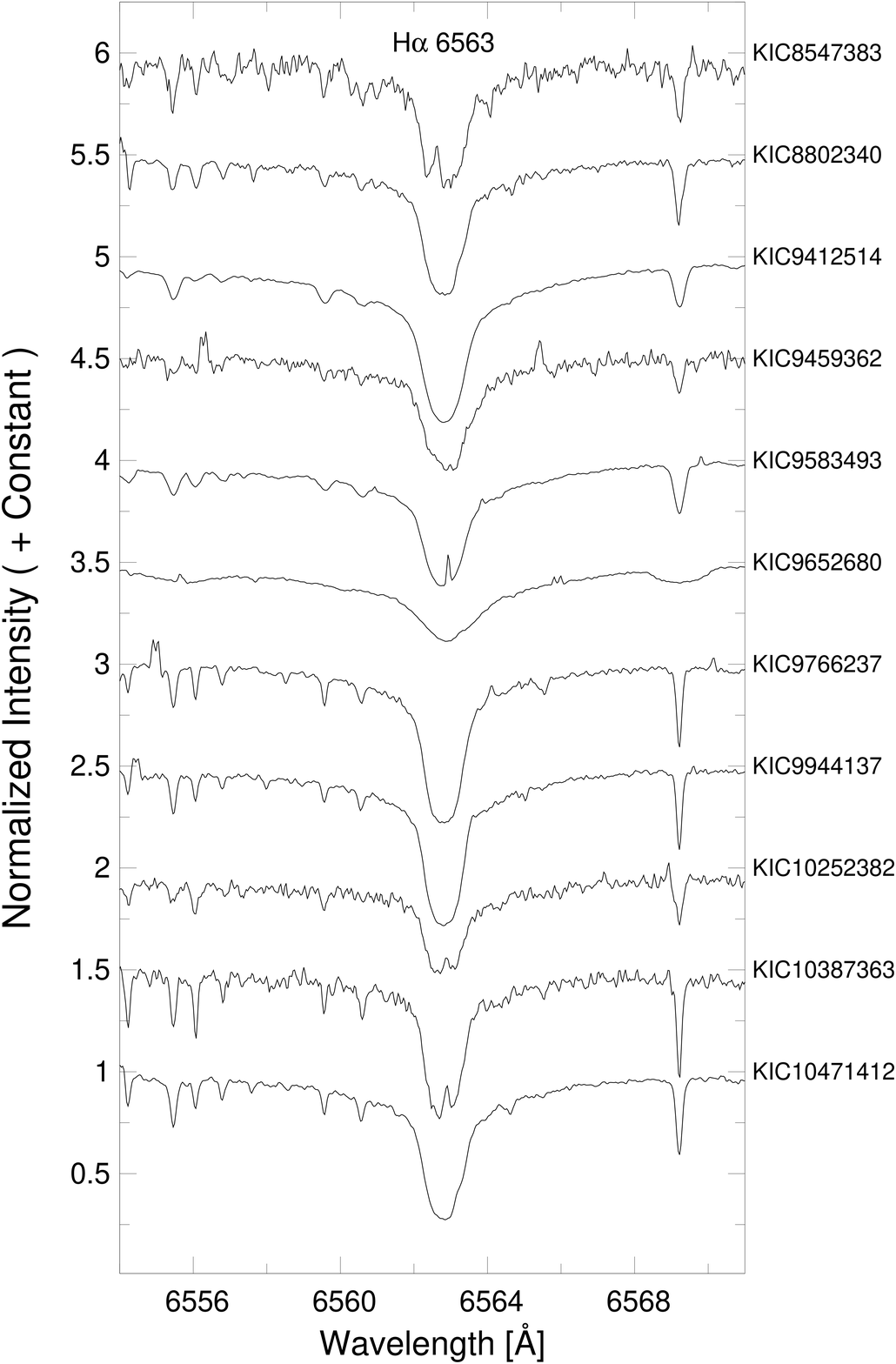}
  \FigureFile(70mm,70mm){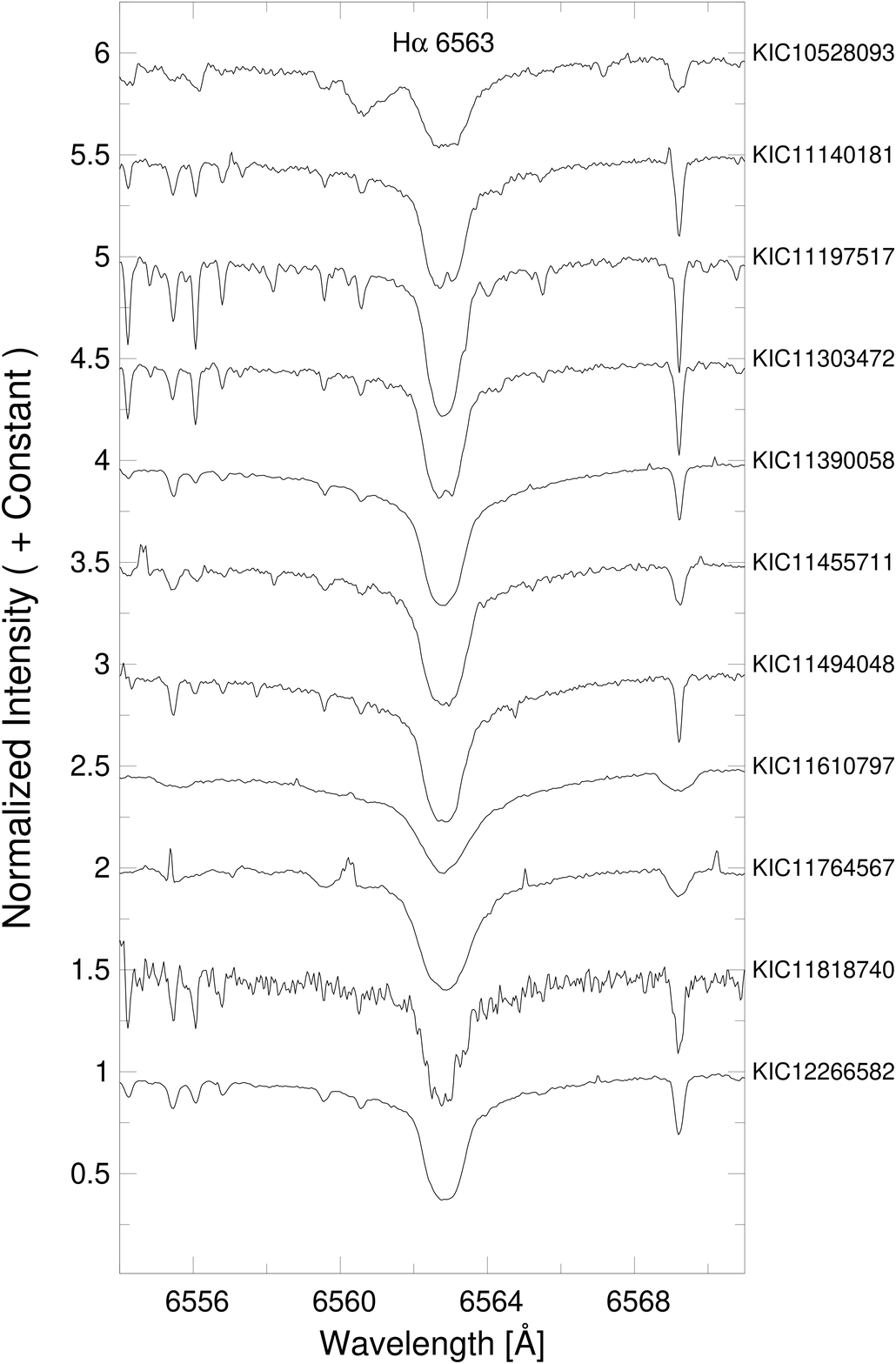}  
   \FigureFile(70mm,70mm){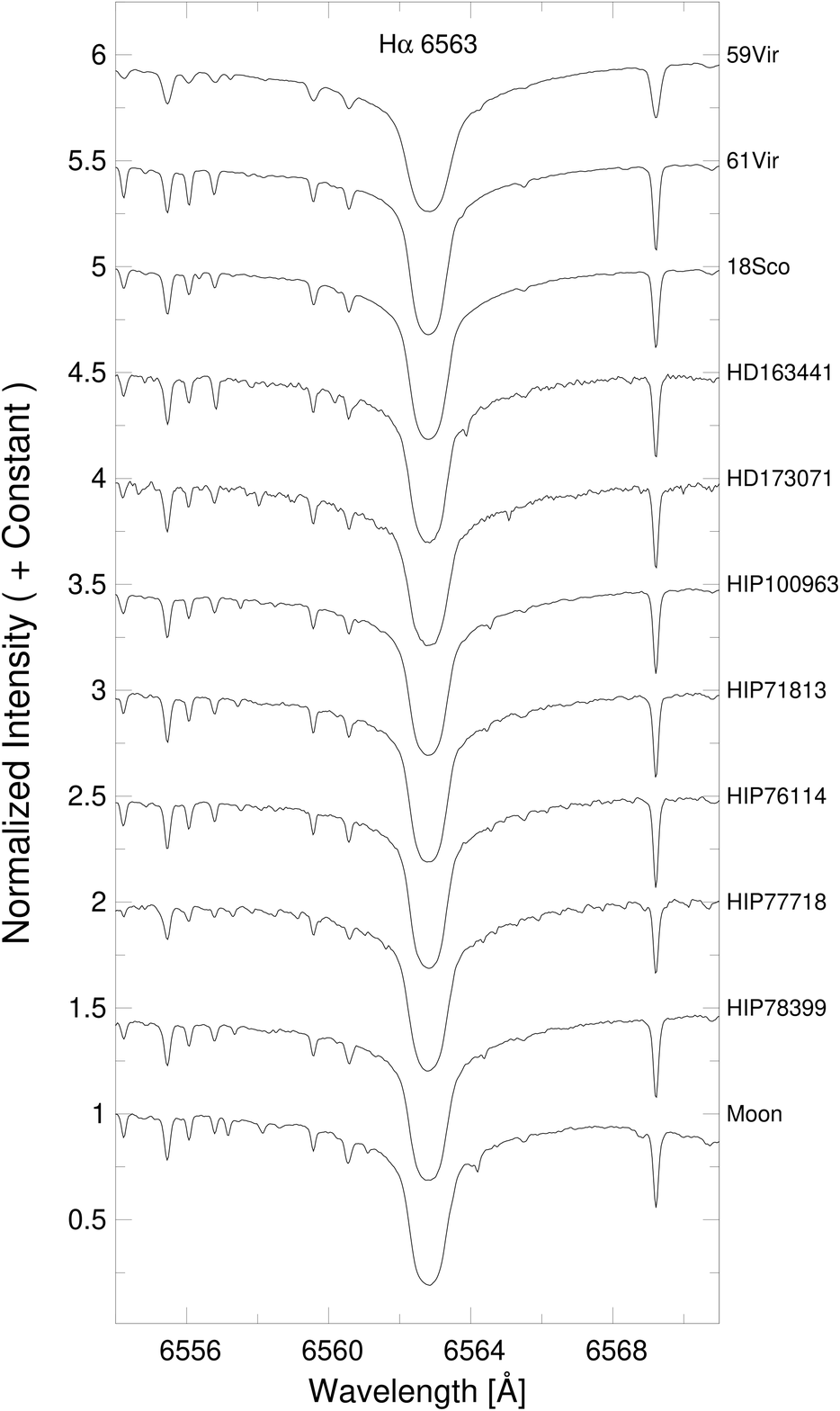}
 \end{center}
\caption{Spectra around H$\alpha$ 6563 line of the 34 superflare stars that show no evidence of binarity, 10 comparison stars, and Moon.
The wavelength scale is adjusted to the laboratory frame. Co-added spectra are used here in case that the star was observed multiple times.
}\label{fig:sg1Ha}
\end{figure}

\begin{figure}[htbp]
 \begin{center}
  \FigureFile(70mm,70mm){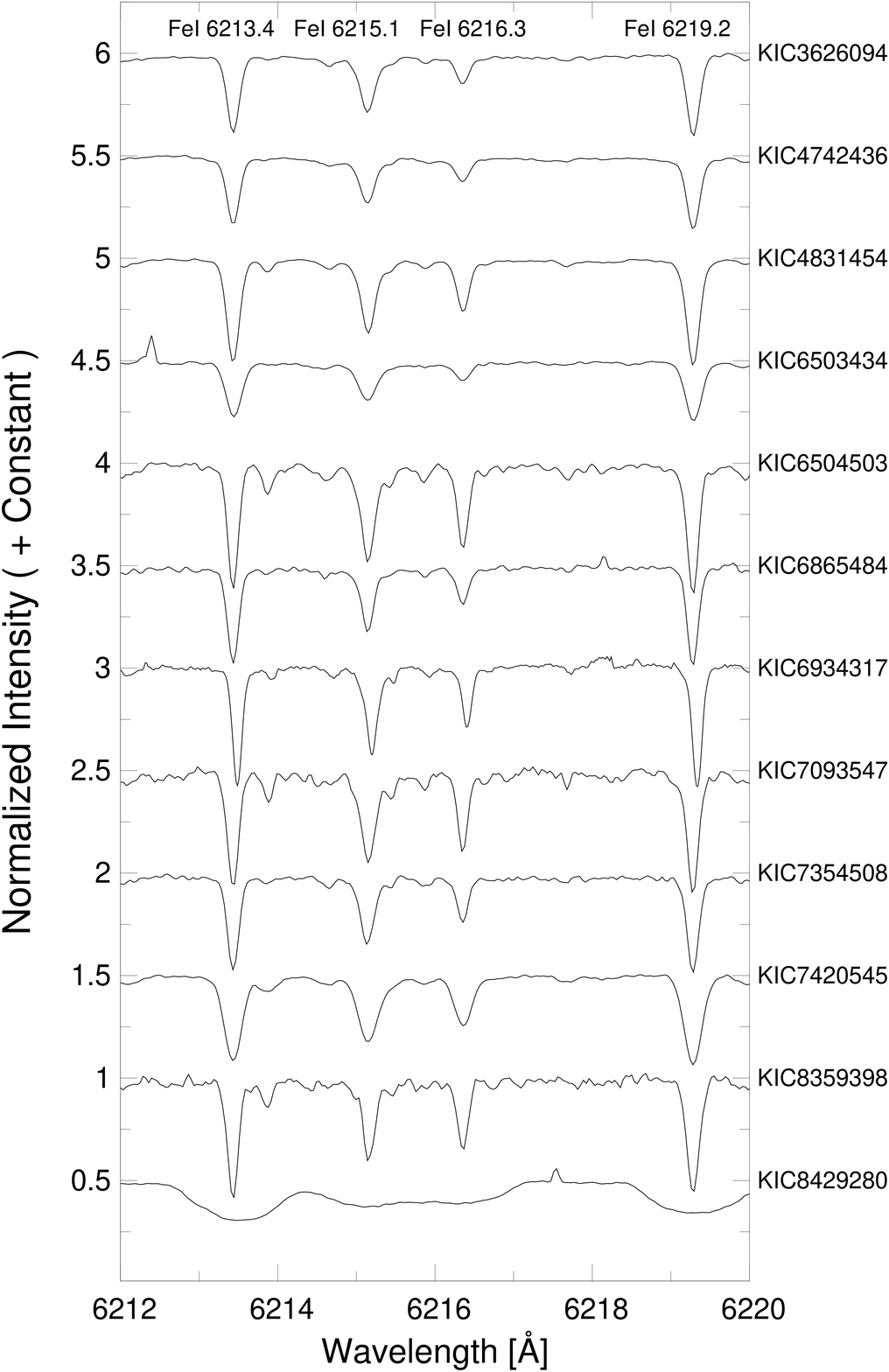}
  \FigureFile(70mm,70mm){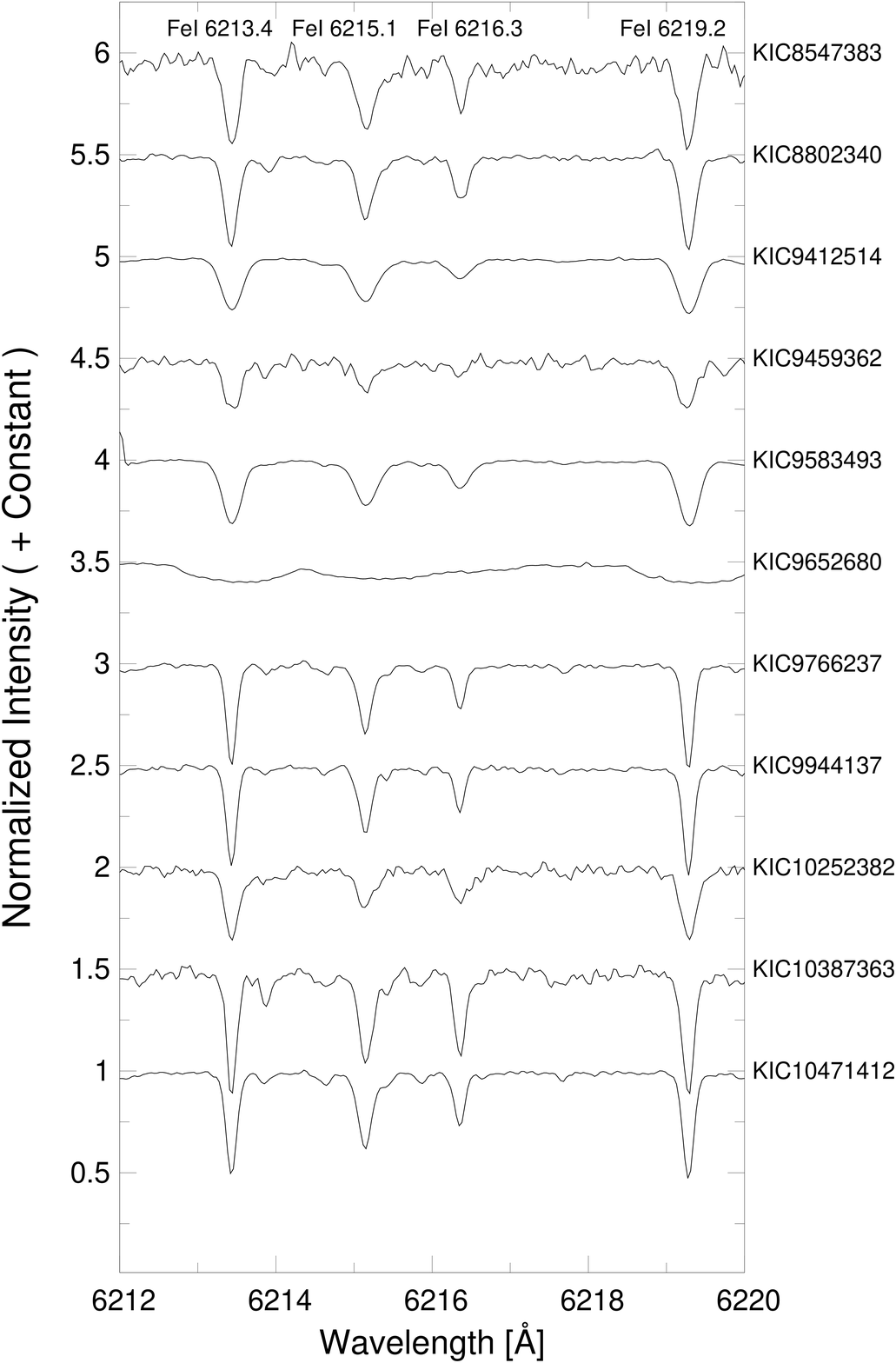}
  \FigureFile(70mm,70mm){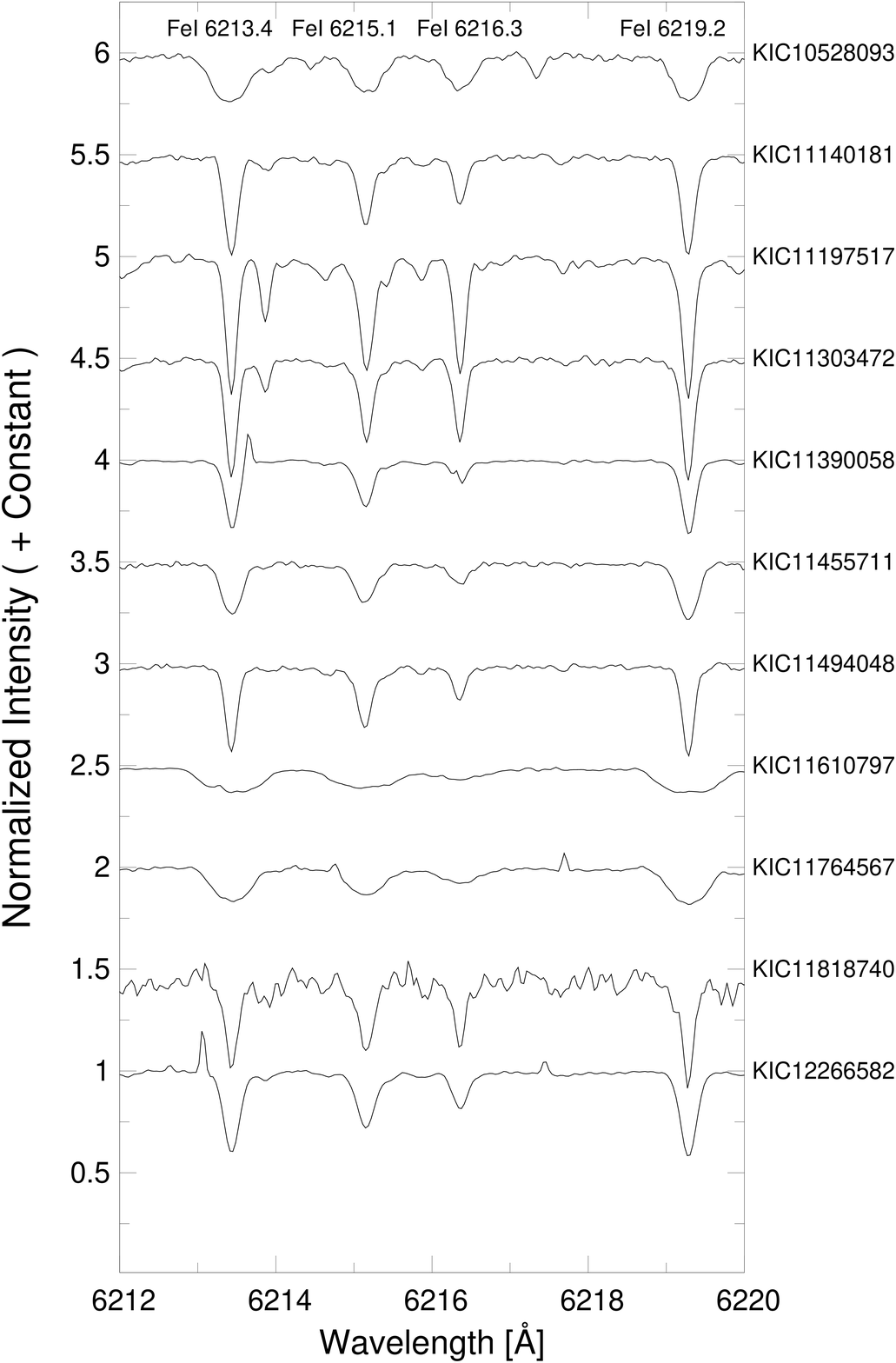}
  \FigureFile(70mm,70mm){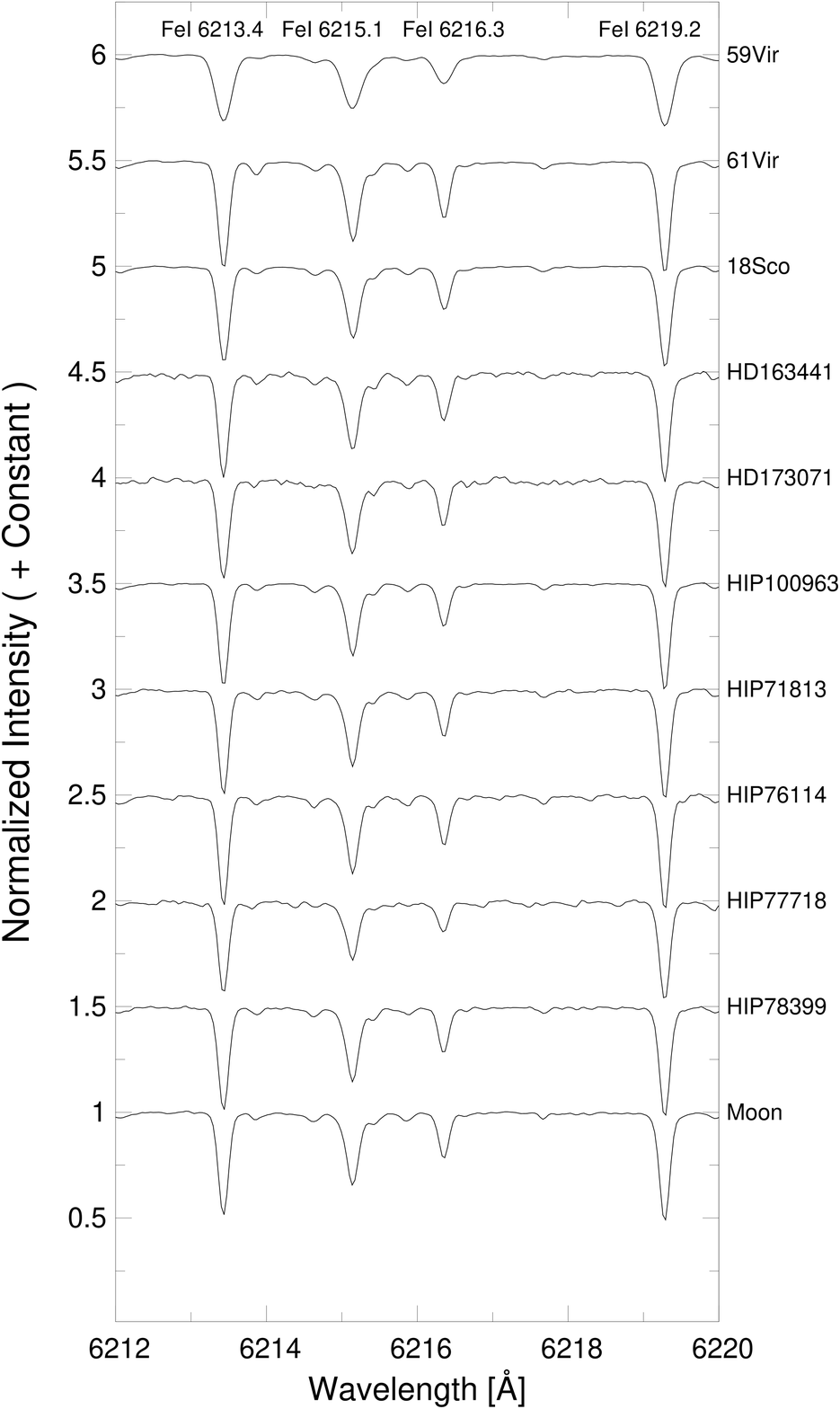}
 \end{center}
\caption{Example of photospheric absorption lines, including Fe I 6213, 6215, 6216, and 6219, of the 34 superflare stars that show no evidence of binarity, 10 comparison stars, and Moon. 
The wavelength scale is adjusted to the laboratory frame. Co-added spectra are used here in case that the star was observed multiple times.}\label{fig:sg1Fe}
\end{figure}

The details of our spectroscopic observations are described in Section 2 of Paper I.
Our observations have been performed by using the High Dispersion Spectrograph (HDS: \cite{Noguchi2002}) at the 8.2-m Subaru telescope.
The names of the above 34 superflare stars, which show no evidence of binarity in Paper I, and 10 comparison stars are shown 
in Table \ref{tab:r0CaHa}.
In addition to the above 10 comparison stars, 
Moon was also observed as a comparison star in this observation, as briefly mentioned in Section 2.2 of Paper I.
The observed spectra of these stars around Ca II 8542 and H$\alpha$ 6563 
are shown in Figures \ref{fig:sg1Ca} and \ref{fig:sg1Ha}, respectively. 
We did not use Ca II 8542 line of 7 target superflare stars 
(KIC7093547, KIC8359398, KIC8547383, KIC9459362, KIC10252382, KIC10387363, and KIC11818740) 
since the resultant S/N ratio around Ca II IRT lines is low, as we can see in Figure \ref{fig:sg1Ca}.
We show these stars with a comment ``low S/N" in this figure.
Examples of spectra around photospheric lines, including Fe I 6212, 6215, 6216, 6219 are shown in Figure \ref{fig:sg1Fe}. 
As we have already mentioned in Paper I,
we observed 13 superflare stars among these 34 superflare stars, and 3 comparison stars (59 Vir, 61 Vir, and 18 Sco) multiple times.
We made co-added spectra of these 16 stars in total, which have ``S12B-comb" or ``S13A-comb" in Supplementary Table 2 of Paper I.
In Figures \ref{fig:sg1Ca}, \ref{fig:sg1Ha}, and \ref{fig:sg1Fe}, co-added spectra of these 16 stars are used. 
Only the co-added spectra are used in this paper when we analyze the spectral data of these 16 stars.
\\
\\
\section{Measurements of stellar activity indicators}\label{sec:anaII}
\subsection{Measurements of Ca II Infrared Triplet and H$\alpha$}\label{subsec:CaIIHa}
In order to investigate the chromospheric activity of the target stars, 
we measured $r_{0}$(8498), $r_{0}$(8542), $r_{0}$(8662), and $r_{0}$(H$\alpha$) index, 
which are the residual core flux normalized by the continuum level 
at the line cores of the Ca II IRT and H$\alpha$, respectively.
As we have already introduced previous researches in detail in Section 3.3 of \citet{SNotsu2013}, 
these indexes are known to be good indicators of stellar chromospheric activity (e.g., \cite{Linsky1979}; \cite{Takeda2010}).
As the chromospheric activity is enhanced, the intensity of these indicators becomes large since a greater amount of emission from the chromosphere fills in the core of the lines. 
The values of $r_{0}$(8498), $r_{0}$(8542), $r_{0}$(8662), and $r_{0}$(H$\alpha$) indexes of the target single superflare stars and comparison stars are listed 
in Table \ref{tab:r0CaHa}. 
We were not able to measure the values of Ca II IRT indexes ($r_{0}$(8498), $r_{0}$(8542), and $r_{0}$(8662)) of 7 target superflare stars 
(KIC7093547, KIC8359398, KIC8547383, KIC9459362, KIC10252382, KIC10387363, and KIC11818740) 
because of low S/N ratios around Ca II IRT lines. 
For reference, we also measured the emission flux of Ca II IRT and H$\alpha$ lines in Appendix \ref{sec:excessFlux}. 
\\ 
\\ 
\subsection{Indirenct estimation of mean intensity of the stellar magnetic field using Ca II 8542 line.}\label{subsec:Hida}
\begin{figure}[htbp]
 \begin{center}
  \FigureFile(120mm,120mm){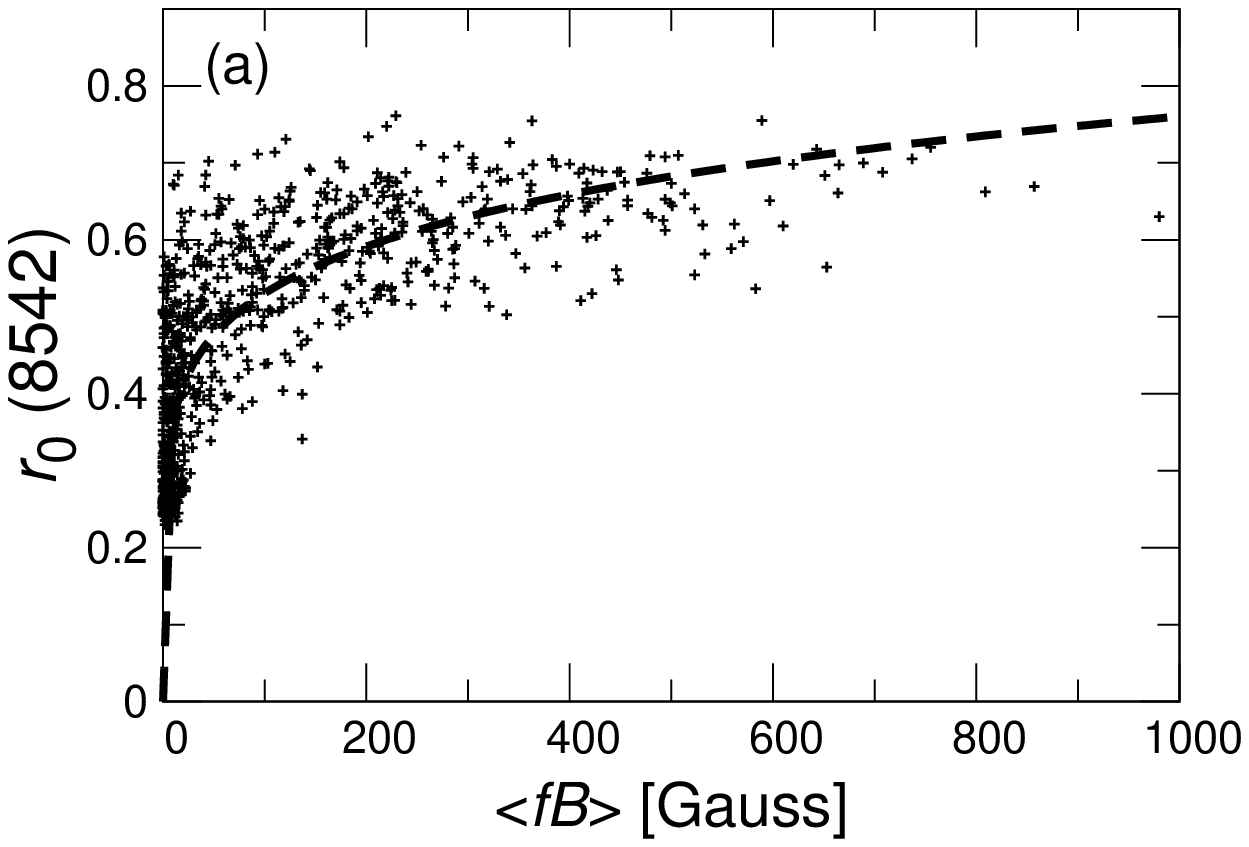}
  \FigureFile(120mm,120mm){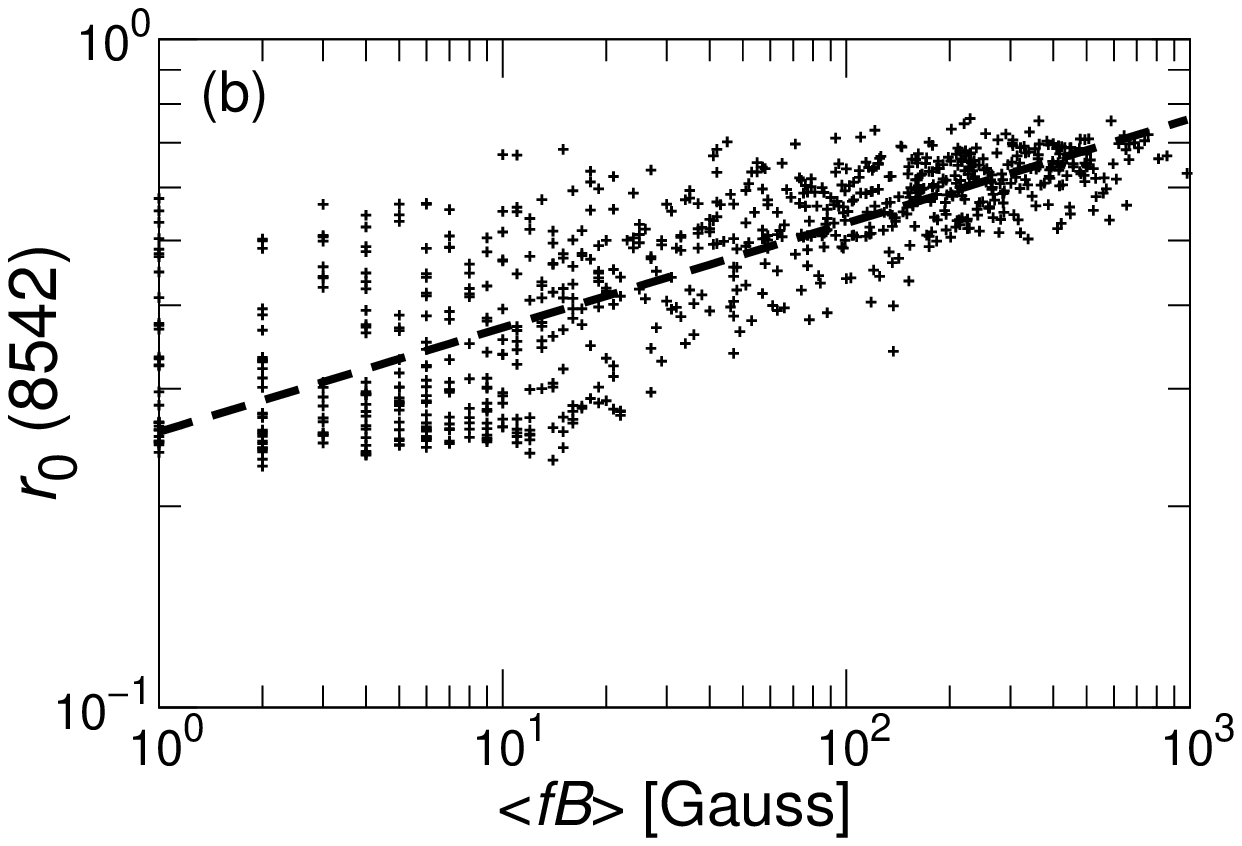}
 \end{center}
\caption{(a) Residual flux normalized by the continuum level at the core of Ca II 8542 line ($r_{0}$(8542) index) vs.
the intensity of photospheric magnetic field ($\langle fB\rangle$), on the basis of the data of our solar observation using DST, Hida Observatory.
The dashed line corresponds to Equation (\ref{eq:r0fB}). \\
(b) Log-log plot of the data shown in (a). The dashed line also corresponds to Equation (\ref{eq:r0fB}).
}\label{fig:Hida-r0fB}
\end{figure}

As mentioned above, Ca II lines are good indicators of stellar chromospheric activity, 
and the intensity of them well correlates with the intensity of stellar magnetic field. 
\citet{Schrijver1989} derived a rough empirical relation between the emission core flux of Ca II K line and the average photospheric magnetic field, 
on the basis of spectroheliographic observations of solar active regions. 
In this paper, we extended this relation to the emission core flux of Ca II 8542 ($r_{0}$(8542)) and the average stellar magnetic field
by newly carrying out spectroheliographic observations of solar active regions.
We summarized the method and results in the following of this section.
\\ \\
\ \ \ \ \ \ \
In order to derive an empirical relation, we compared $r_{0}$(8542) values 
obtained from spectroheliographic observation of the active region NOAA11654, with the photospheric magnetic field strength data of the same region on the same observation date.
The spectroheliographic observation was conducted using 60cm Domeless Solar Telescope (DST) 
with the horizontal spectrograph of the Hida Observatory, Kyoto University \citep{Nakai1985}. 
The observation date was 2013 January 11th (JST). 
The photospheric magnetic field strength data ($\langle fB\rangle$; ``$f$" is a filling factor) that we used here were taken by HMI (\cite{Scherrer2012}; \cite{Schou2012}) 
on the Solar Dynamics Observatory (SDO; \cite{Pesnell2012}) on the same day.
\\ \\
\ \ \ \ \ \ \ 
Adjusting the spatial scale of both the DST and HMI data to $2".4\times 2".4$, 
we plotted $r_{0}$(8542) value of each data point in the target active region as a fuction of $\langle fB\rangle$ (mean intensity of the photospheric magnetic field) 
of the same point.
As shown in Figure \ref{fig:Hida-r0fB} (a), a rough positive correlation is seen between $r_{0}$(8542) and $\langle fB\rangle$.
This figure means that $r_{0}$(8542) index roughly reflects the average intensity of the stellar magnetic field ($\langle fB\rangle$), 
though this correlation is not so good especially for less active stars (low $\langle fB\rangle$). 
Using the least-square method, we conduct a linear fit in the double logarithmic graph $\log r_{0}$(8542) vs. $\log (\langle fB\rangle)$ (Figure \ref{fig:Hida-r0fB} (b)).
As a result, we derived a very rough empirical relation between $r_{0}$(8542) and $\langle fB\rangle$: 
\begin{equation}\label{eq:r0fB}
\log r_{0}(8542)=-0.587+0.156\times\log (\langle fB\rangle) \ .
\end{equation}
\ \ \ \ \ \ \
We then roughly estimate $\langle fB\rangle$ values of the target stars, 
by using Equation (\ref{eq:r0fB}) and $r_{0}$ (8542) values.
The estimated $\langle fB\rangle$ values are listed in Table \ref{tab:r0CaHa}.
The error value of each $\langle fB\rangle$ value in Table \ref{tab:r0CaHa} is estimated from the standard deviation of $\langle fB\rangle$ distribution 
as a function of $r_{0}$(8542) in Figure \ref{fig:Hida-r0fB}.
In addition, if we apply Equation (\ref{eq:r0fB}) to solar $r_{0}$(8542) value estimated in \citet{Takeda2010} ($r_{0}$(8542)=0.193), 
$\langle fB\rangle$ is estimated to be $0.2\pm6$ [Gauss]. 
This value is consistent with the values of solar mean magnetic field (a few Gauss) within its error range, though the error range is large.
\\ 
\\ 
\section{Rotational Velocity and Inclination Angle}\label{subsec:inc}
\begin{figure}[htbp]
 \begin{center}
  \FigureFile(150mm,150mm){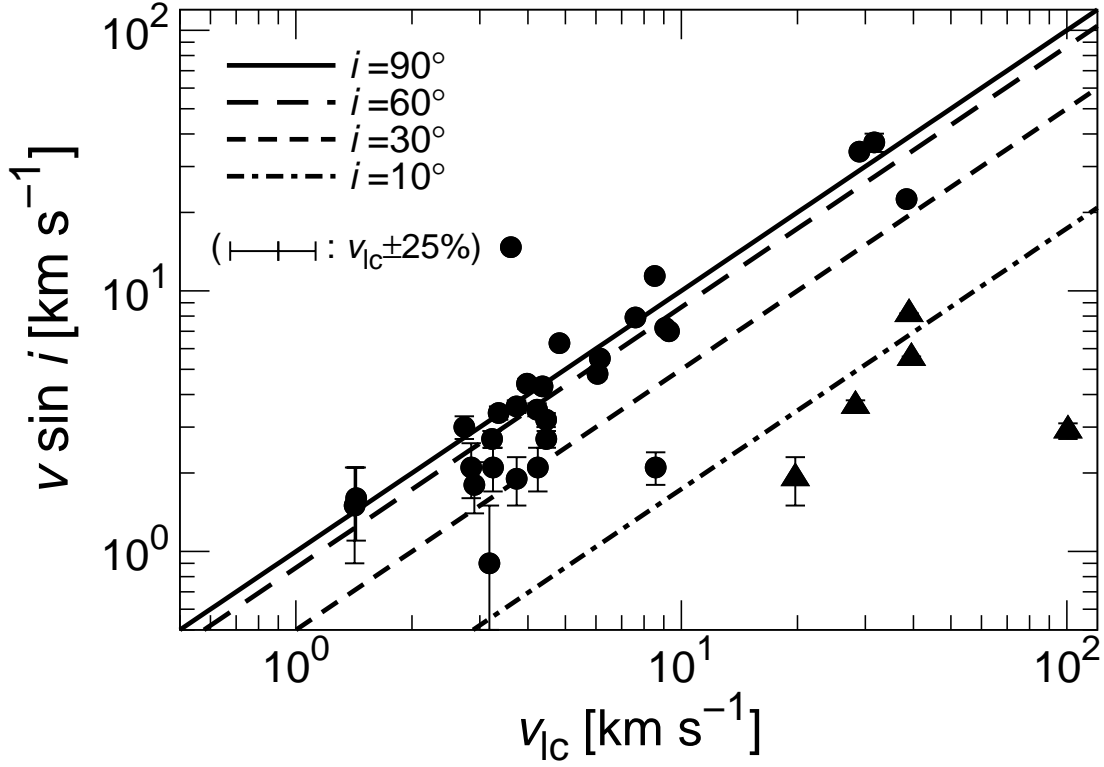}
 \end{center}
\caption{
Projected rotational velocity ($v \sin i$) as a function of the stellar rotational velocity ($v_{\rm{lc}}$) estimated from 
the period of the brightness variation and stellar radius. 
The typical error of $v_{\rm{lc}}$ is about $\pm$25\%~of each value.
The error value of $v \sin i$ is listed in Table 4 of Paper I.
The solid line represents the case that our line of sight is vertical to the stellar rotation axis ($i=90^{\circ}$; $v \sin i=v_{\rm{lc}}$).
We also plot three different lines, which correspond to smaller inclination angles ($i=60^{\circ}$, $30^{\circ}$, $10^{\circ}$).
Filled triangles represent superflare stars whose inclination angle is especially small ($i\leq 13^{\circ}$), 
while filled circles represent the other stars ($i>13^{\circ}$).
}\label{fig:vlc-vsini}
\end{figure}

\begin{figure}[htbp]
 \begin{center}
  \FigureFile(130mm,130mm){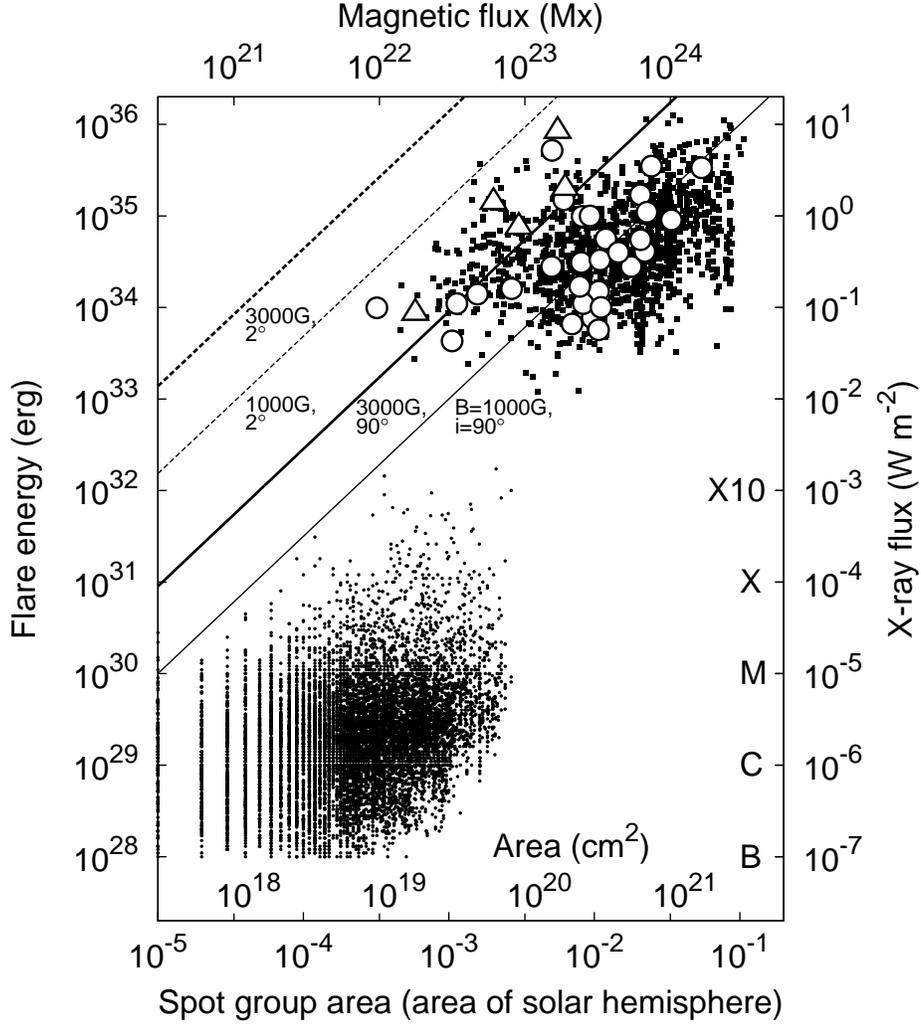}
 \end{center}
\caption{
Scatter plot of the flare energy as a function of the spot coverage.
The data of superflares on solar-type stars (filled squares) and solar flares (filled circles) in this figure are 
completely the same as those in Figure 10 of \citet{YNotsu2013}.
Thick and thin solid lines corresponds to the analytic relation 
between the spot coverage and the flare energy, which is obtained from Equation (14) of \citet{YNotsu2013} for $B$=3,000G and 1,000G. 
The thick and thin dashed lines correspond to the same relation in case of $i=2^{\circ}$ (nearly pole-on) for $B$=3,000G and 1,000G. 
The right-hand vertical axis (X-ray flux) and the horizontal axis at the top (Magnetic flux) are roughly estimated 
in the completely same way as done in Figure 10 of \citet{YNotsu2013}.
Open circle and triangle points on the filled squares represent the data points of the most energetic superflare event reported in \citet{Shibayama2013}
of the 34 target superflare stars. Their energy values are listed in 9th column of Table 1 of Paper I.
Open triangles represent superflare stars whose inclination angle is especially small ($i\leq 13^{\circ}$), 
while open circles represent the others ($i>13^{\circ}$). This classification is on the basis of Figure \ref{fig:vlc-vsini}.
}\label{fig:spotene}
\end{figure}

As mentioned above, we reported the values of projected rotational velocity ($v \sin i$), stellar radius ($R_{\rm{s}}$), 
and the brightness variation period ($P_{0}$) of the 34 ``single" target superflare stars in Paper I.
The values of $v \sin i$ and $P_{0}$ are listed again in Table \ref{tab:r0CaHa}.
Assuming that the brightness variations of these stars are caused by the rotation
of the stars with starspots, we can estimate the rotational velocity ($v_{\rm{lc}}$) from $P_{0}$ and $R_{\rm{s}}$ by using 
\begin{equation}\label{eq:vlc}
v_{\rm{lc}}=\frac{2\pi R_{\rm{s}}}{P_{0}} \ ,
\end{equation}
as we have already mentioned in Section 5.2 of \citet{SNotsu2013}. 
The resultant values of $v_{\rm{lc}}$ are listed in Table \ref{tab:r0CaHa}.
We here consider that the typical error of $v_{\rm{lc}}$ is about $\pm$25\%, 
assuming that the upper limit of errors of $R_{\rm{s}}$ and $P_{0}$ are about 20\% and 10\%, respectively, on the basis of Paper I. 
\\ \\
\ \ \ \ \ \ \
In Figure \ref{fig:vlc-vsini}, we plot $v \sin i$ as a function of the $v_{\rm{lc}}$. 
Some data points in Figure \ref{fig:vlc-vsini} show differences between the values of $v_{\rm{lc}}$ and $v \sin i$.
The projected rotational velocity ($v \sin i$) tends to be smaller than $v_{\rm{lc}}$.
Such differences should be explained by the inclination effect, as mentioned in the previous studies (e.g., \cite{Hirano2012}; \cite{SNotsu2013}).
On the basis of $v \sin i$ and $v_{\rm{lc}}$, the stellar inclination angle ($i$) can be estimated by using the following relation:
\begin{equation}\label{eq:inc}
i=\arcsin\Biggl(\frac{v \sin i}{v_{\rm{lc}}}\Biggr) \ .
\end{equation}
In Figure \ref{fig:vlc-vsini}, we also show four lines indicating $i=90^{\circ}$ ($v \sin i=v_{\rm{lc}}$), $i=60^{\circ}$, $i=30^{\circ}$, and $i=10^{\circ}$.
This figure shows two following important results. 
First,  for almost all the stars (33 stars) except for KIC11764567, the relation ``$v \sin i\lesssim v_{\rm{lc}}$" is satisfied.
This is consistent with our assumption that the brightness variation is caused by the rotation 
since the inclination effect mentioned above can cause the relation ``$v \sin i\lesssim v_{\rm{lc}}$" if 
$v_{\rm{lc}}$ values really correspond to the rotational velocities (i.e. $v=v_{\rm{lc}}$).
This is also supported by another fact that the distribution of the data points in Figure \ref{fig:vlc-vsini} are not random. 
Their distribution is expected to be much more random if the brightness variations have no relations with the stellar rotation.
Second, stars that are distributed in the lower right side of Figure \ref{fig:vlc-vsini} are expected to have small inclination angles and to be nearly pole-on stars.
In this figure, we distinguish such five stars with especially small inclination angle ($i\leq13^{\circ}$) (KIC3626094, KIC4742436, KIC6503434, KIC6934317, and KIC9412514) 
from the other stars, using filled triangle data points. 
\\ \\
\ \ \ \ \ \ \
We can confirm the above inclination effects from another point of view, as already mentioned for KIC6934317 in Figure 7 of \citet{SNotsu2013}.
Figure \ref{fig:spotene} is a scatter plot of the flare energy of superflares and solar flares as a function of the spot coverage.
The data used in this figure are the same as in Figure 10 of \citet{YNotsu2013}. 
The spot coverage of superflare stars are calculated from the amplitude of stellar brightness variations, 
and the energy of superflares are estimated by using the superflare amplitude and the duration time \citep{Shibayama2013}.
Thick and thin solid lines correspond to the analytic relation 
between the spot coverage and the flare energy, which is obtained from Equation (14) of \citet{YNotsu2013} in case of $i=90^{\circ}$ for $B$=3,000G and 1,000G. 
The thick and thin dashed lines correspond to the same relation in case of $i=2^{\circ}$ (nearly pole-on) for $B$=3,000G and 1,000G, 
assuming that the brightness variation becomes small as a result of the inclination effect.
These lines are considered to give an upper limit of superflare energy for each inclination angle \citep{YNotsu2013}.
Considering these things, the superflare stars located in the upper left side of this figure are expected to have a low inclination angle.
Open triangles in Figure \ref{fig:spotene} correspond to superflare stars whose inclination angle is especially small ($i\leq 13^{\circ}$) on the basis of Figure \ref{fig:vlc-vsini}, 
while open circles represent the other stars ($i>13^{\circ}$). This classification is the same as that in Figure \ref{fig:vlc-vsini}.
All of the five triangle data points are located above the thick solid line ($i=90^{\circ}$ and $B$=3000G). 
This means that these stars are also confirmed to have low inclination angle on the basis of Figure \ref{fig:vlc-vsini}.
As a result, these two figures (Figure \ref{fig:vlc-vsini} and \ref{fig:spotene}) are confirmed to be consistent, 
and in other words, the stellar projected rotational velocity spectroscopically measured is consistent with the rotational velocity estimated from the brightness variation.
The fact indicates that the brightness variation of superflare stars is caused by the rotation.
\\ 
\\
\section{Stellar Chromospheric Activity and Starspots of Superflare Stars}\label{subsec:activity-spots}
\begin{figure}[htbp]
 \begin{center}
  \FigureFile(100mm,100mm){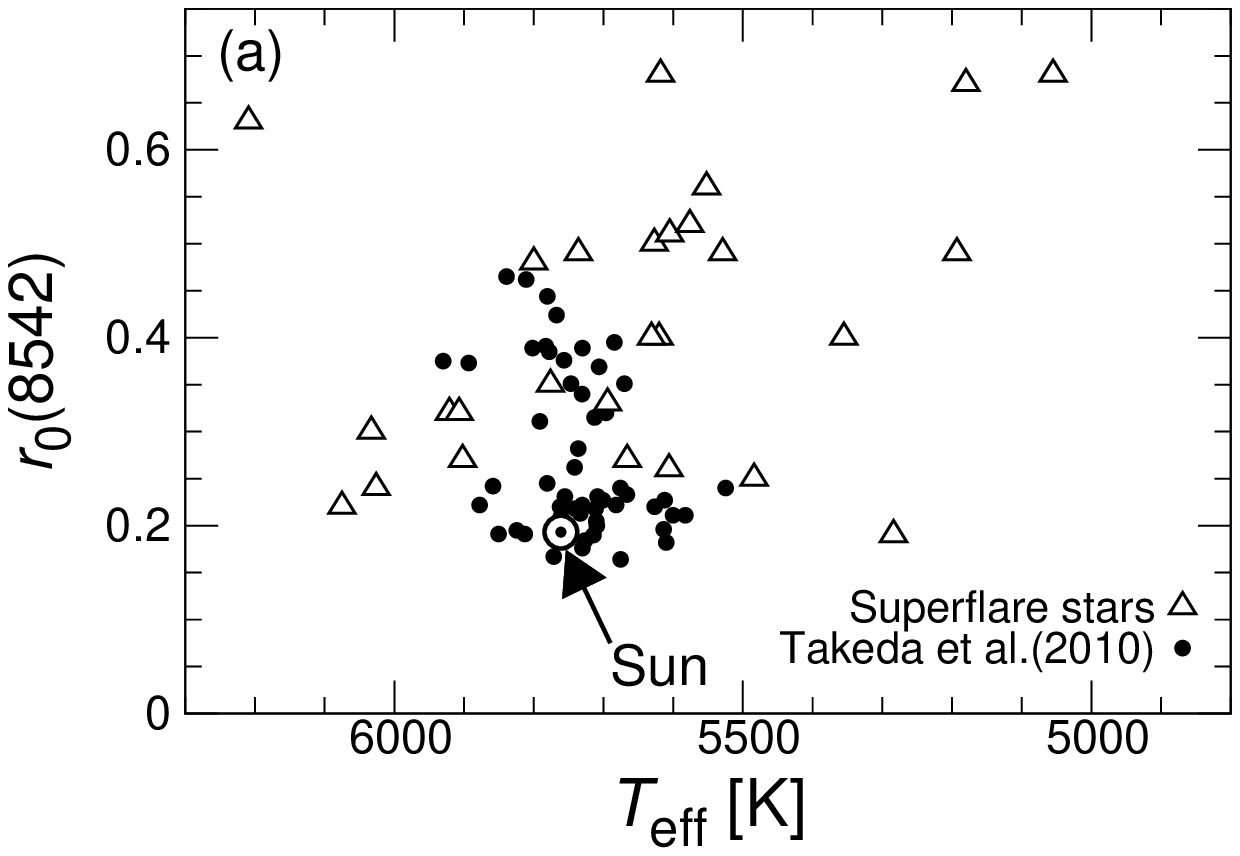}
   \FigureFile(100mm,100mm){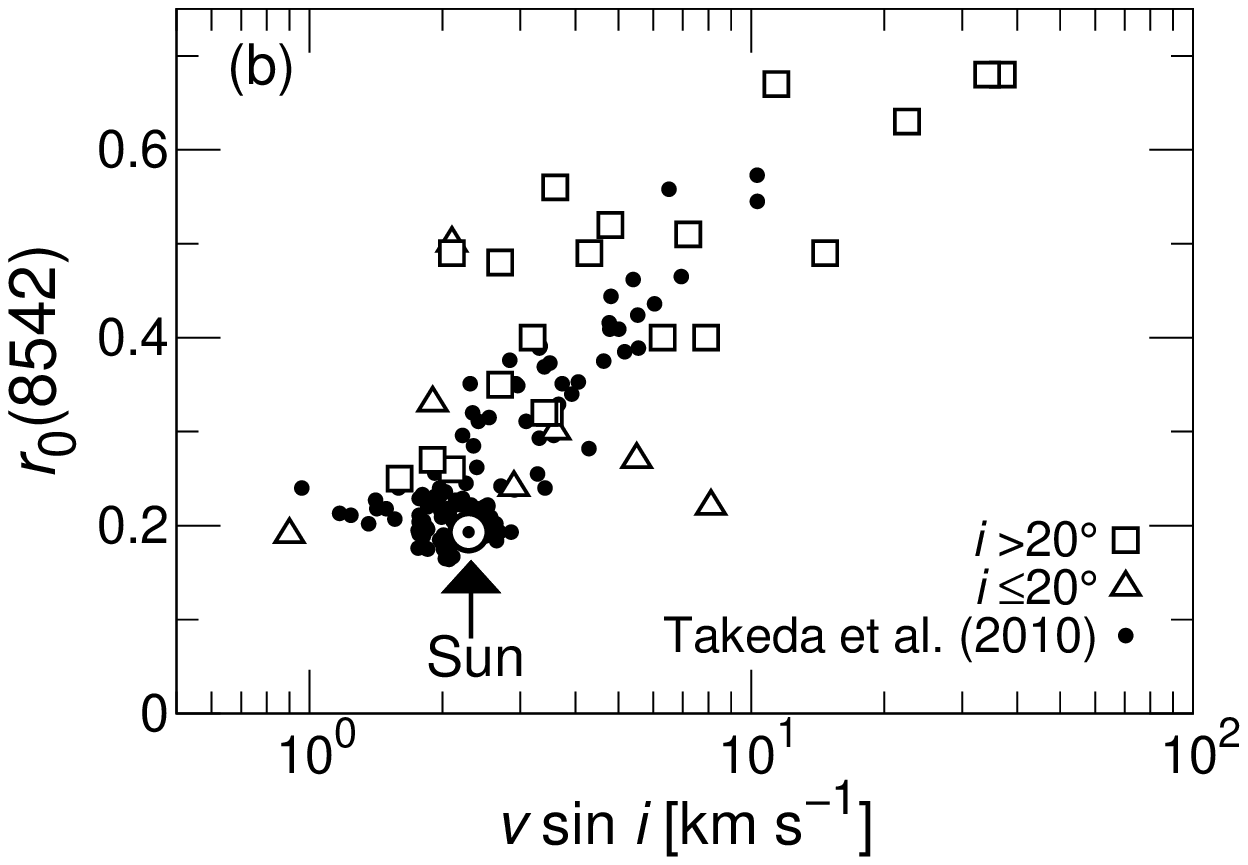}
 \end{center}
\caption{
(a) $r_{0}$(8542) as a function of $T_{\rm{eff}}$ of the stars.
In addition to the results of the target superflare stars (open triangles), 
we also plotted the data of ordinary solar-type stars in \citet{Takeda2010} (black filled circles). 
The solar $r_{0}$(8542) value in \citet{Takeda2010} ($r_{0}(8542)=0.193$ and $T_{\rm{eff}}=5761$K) is also plotted using a circled dot point for reference.
\\
(b) $r_{0}$(8542) as a function of $v\sin i$.
The results of the target superflare stars are plotted, being classified into two groups 
on the basis of stellar inclination angles ($i$) estimated in Section \ref{subsec:inc}.
Open squares represent superflare stars with $i>20^{\circ}$, while open triangles represent those with $i\leq20^{\circ}$.
The data of ordinary solar-type stars reported in \citet{Takeda2010} are plotted by using black circle points. 
The solar $r_{0}$(8542) and $v \sin i$ value in \citet{Takeda2010} ($r_{0}(8542)=0.193$ and $v \sin i=2.29$ km s$^{-1}$) is also plotted 
by using a circled dot point for reference.
}\label{fig:Tvr0}
\end{figure}

\begin{figure}[htbp]
 \begin{center}
  \FigureFile(105mm,105mm){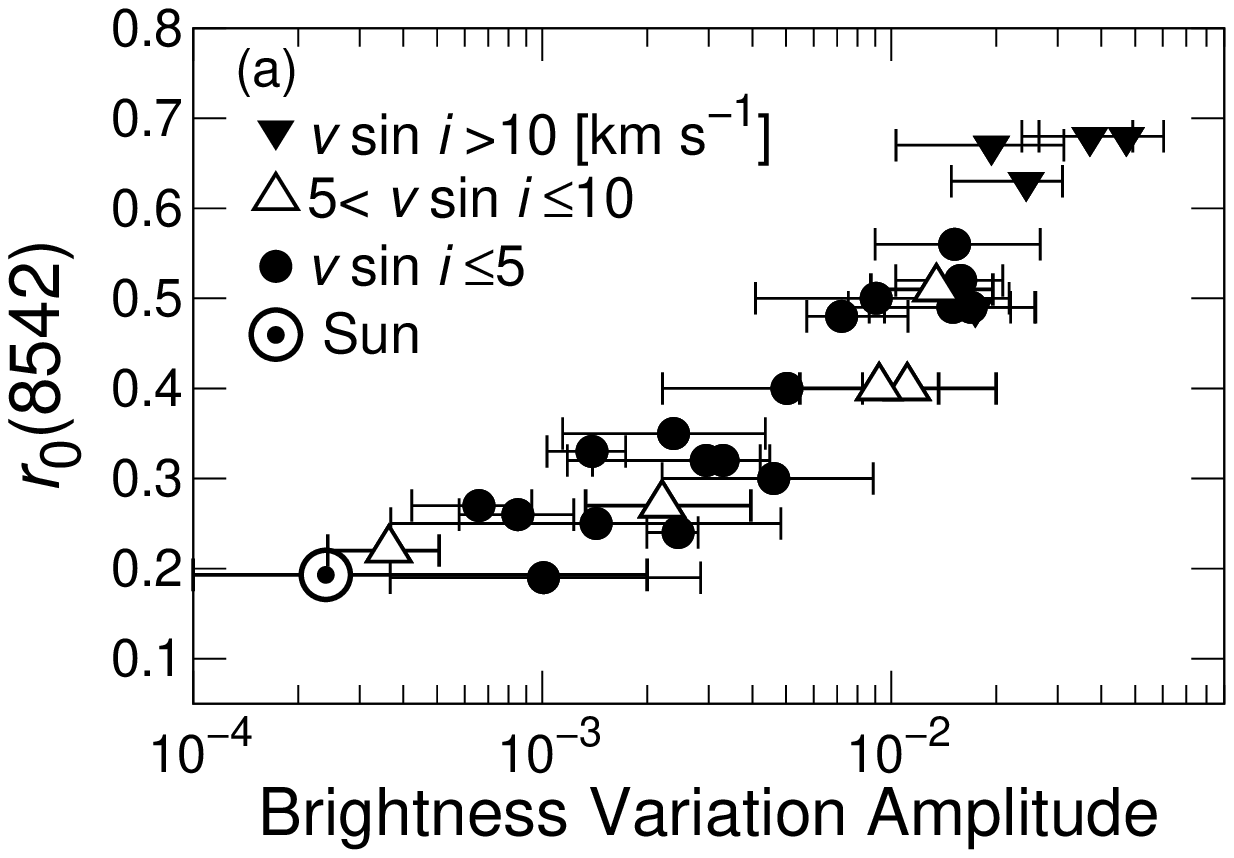} 
  \FigureFile(105mm,105mm){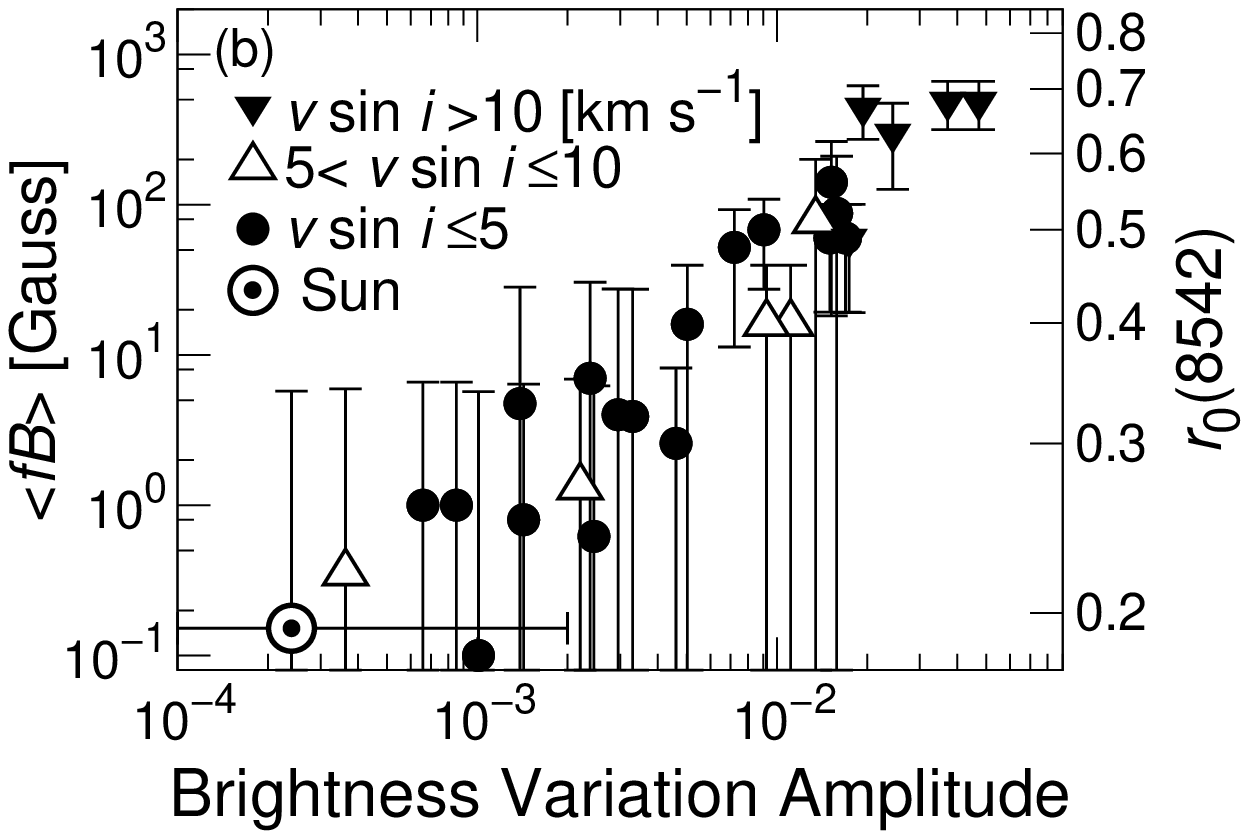}
 \end{center}
\caption{(a) $r_{0}$(8542) as a function of
the amplitude of stellar brightness variation that are listed as $\langle$BVAmp$\rangle$ in Table \ref{tab:r0CaHa}.
The results of the target superflare stars are plotted, being classified into three groups 
on the basis projected rotational velocity.
Black filled inverted triangles represent superflare stars with $v \sin i>10$ km s$^{-1}$, 
open regular triangles represent those with $5<v \sin i\leq 10$ km s$^{-1}$, 
and filled circles represent those with $v \sin i\leq5$ km s$^{-1}$.
The solar value is plotted by using a circled dot point.
\\
(b) $\langle fB\rangle$ as a function of amplitude of stellar brightness variation.
The results of the target superflare stars are plotted, being classified into three groups using the same symbols as in (a).
The errorbars of $\langle$BVAmp$\rangle$ are not plotted here 
since it is hard to see if vertical and horizontal error bars are both plotted.
The $r_{0}$(8542) values shown on the right-hand vertical axis correspond to $\langle fB\rangle$ on the left-hand vertical axis on the basis of Equation (\ref{eq:r0fB}). 
The solar value is also plotted by using a circled dot point.
}\label{fig:ampr0fB}
\end{figure}

In Figure \ref{fig:Tvr0} (a), we plot $r_{0}$(8542) as a function of $T_{\rm{eff}}$ of the target superflare stars. 
The data of ordinary solar-type stars reported in \citet{Takeda2010} are also plotted in this figure.
The variability range of $r_{0}$(8542) between the solar maximum and solar minimum is only 0.19$\sim$0.20 \citep{Livingston2007}.
This figure shows that almost all the target superflare stars are more active compared to the Sun from the viewpoint of the $r_{0}$(8542) index.
In other words, the mean magnetic field strength of the target stars can be higher than that of the Sun (see Section \ref{subsec:Hida}).
In addition, 4 target superflare stars (KIC8429280, KIC9652680, KIC10528093, and KIC11610797) 
show much higher chromospheric activity ($r_{0}$(8542)$>$0.6) compared to the other solar-type stars in this figure.
\\ \\
\ \ \ \ \ \ \
In Figure \ref{fig:Tvr0} (b), we plot $r_{0}$(8542) as a function of $v \sin i$ of the target superflare stars. 
The data of ordinary solar-type stars reported 
in \citet{Takeda2010} \footnote{The dotted line in Figure 5 (a) of \citet{Takeda2010} shows
the expected variation of $r_{0}$(8542) due to the blurring effect caused by an increase of $v \sin i$ is 
small compared to the real changes of the stellar activity level.
Then we here consider that this effect does not have any essential effects 
on the interpretation of Figure \ref{fig:Tvr0} (b).} 
are also plotted for reference in this figure.
There is a rough positive correlation between $r_{0}$(8542) and $v \sin i$, 
and more rapidly rotating stars have higher chromospheric activity.
All of the 5 stars with extremely high $v \sin i$ value ($v \sin i \geq 10$km s$^{-1}$) compared to the Sun ($v \sin i\sim$2 km s$^{-1}$)
show relatively high $r_{0}$(8542) values ($r_{0}(8542)\gtrsim 0.5$) .
This is consistent with many previous observational results of ordinary main-sequence stars (e.g., \cite{Noyes1984}; \cite{Takeda2010}).
In addition, about 6 superflare stars with relatively small $v \sin i$ values ($v \sin i\lesssim 5$ km s$^{-1}$) in this figure 
(KIC4831454, KIC6865484, KIC8802340, KIC11140181, KIC11303472, and KIC12266582) 
also have relatively high chromospheric activity ($r_{0}(8542)\gtrsim 0.5$) 
compared to the data of ordinary solar-type stars taken from \citet{Takeda2010}.
\\ \\
\ \ \ \ \ \ \
The $r_{0}$(8542) values are plotted in Figure \ref{fig:ampr0fB} (a) as a function of the amplitude of stellar brightness variation ($\langle$BVAmp$\rangle$).
We calculated $\langle$BVAmp$\rangle$ listed in Table \ref{tab:r0CaHa} 
by taking the average of the amplitude values of Q2$\sim$Q16 data. 
The amplitude value in each Quarter is available in Supplementary Data of this paper. 
We did not use Q0, Q1, and Q17 data since the duration of these three quarters is short ($\lesssim$30 days) 
compared to those of the other 15 quarters ($\sim$90 days) \citep{Thompson2013}.
The errors of $\langle$BVAmp$\rangle$ in Table \ref{tab:r0CaHa} correspond to the maximum and minimum of the amplitude values of Q2$\sim$Q16 data.
The solar $r_{0}$(8542) value in \citet{Takeda2010} ($r_{0}(8542)=0.193$) is also plotted for reference.
The solar brightness variation amplitude in this figure is calculated from the sunspot coverage in the observation date (2007 February 2) of \citet{Takeda2010}, 
and its maximum and minimum value shown here correspond to the typical sunspot coverage of the solar maximum and minimum, respectively.
In Figure \ref{fig:ampr0fB} (a), there is a rough positive correlation between $r_{0}$(8542) and $\langle$BVAmp$\rangle$.
Assuming that the brightness variation of superflare stars is caused by the rotation of a star with starspots, 
the brightness variation amplitude ($\langle$BVAmp$\rangle$) corresponds to the starspot coverage of these stars. 
Then, we can say that there is a rough positive correlation between the starspot coverage and chromospheric activity level ($r_{0}$(8542)).
This rough correlation shows us that all the target stars expected to have large starspots 
on the basis of their large amplitude of the brightness variation show high magnetic activity compared to the Sun.
In other words, our assumption that the amplitude of the brightness variation correspond to the spot coverage is supported, 
since high magnetic activity, which are confirmed by using $r_{0}$(8542) values, are considered to be caused by the existence of large starspots.
\\ \\
\ \ \ \ \ \ \
In Figure \ref{fig:ampr0fB} (b), we also plot $\langle fB\rangle$ values that we estimated from $r_{0}$(8542) in Section \ref{subsec:Hida}, as a function of $\langle$BVAmp$\rangle$.
With this figure, we can confirm the same conclusion as we did with Figure \ref{fig:ampr0fB} (a), 
though the errors of $\langle fB\rangle$ values are a bit large especially for less active stars.
In Appendix \ref{sec:excessFlux}, we also investigated the emission flux of Ca II IRT and H$\alpha$ lines for reference, 
and confirmed basically the same conclusions as we did with $r_{0}$(8542) index in this section.
\\
\\
\ \ \ \ \ \ \
In addition to the analyses in this paper, we plan to discuss the Li abundances and stellar age of superflare stars 
in another future paper (Honda et al. in preparation).
As we have introduced previous researches in Section 4.5 of \citet{SNotsu2013}, 
Li abundance can provide some constraints on the age of G-type stars (e.g., \cite{Soderblom1993a}; \cite{Sestito2005}).
Information of stellar age is important since the stellar activity has a deep relation with the stellar age (e.g., \cite{Soderblom1991}),
but this point is beyond the scope of this paper.
\\
\\
\section{Summary}\label{sec:summary}
We have performed high dispersion spectroscopic observations of 50 superflare stars with Subaru/HDS.
More than half (34 stars) of these 50 target superflare stars show no evidence of binarity, 
and we measured stellar parameters of these 34 stars in our previous study (Paper I).
In this paper, we investigated whether the quasi-periodic brightness variation of superflare stars is explained by the rotation, 
and whether superflare stars have large starspots, by using stellar parameters derived in Paper I.
First,  the value of $v \sin i$
measured from spectroscopic results is consistent with the rotational velocity estimated from the period of the brightness variation.
Next, we measured the intensity of Ca II IRT and H$\alpha$ lines, which are good indicators of stellar chromospheric activity, 
and compared them with other stellar properties. 
The intensity of Ca II IRT lines indicates that the mean magnetic field strength ($\langle fB\rangle$) 
of the target superflare stars can be higher than that of the Sun.
We found a correlation between the amplitude of the brightness variation and 
the intensity of Ca II IRT. 
All the targets expected to have large starspots because of their large amplitude of the brightness variation show high 
chromospheric activity compared to the Sun.
These results support that the brightness variation of superflare stars is explained by the rotation with large starspots.  
\\
\\
\\
\\
\bigskip

This study is based on observational data collected with Subaru Telescope, 
which is operated by the National Astronomical Observatory of Japan. 
We are grateful to Dr. Akito Tajitsu and other staffs of the Subaru Telescope 
for making large contributions in carrying out our observations. 
We would also like to thank Dr. Yoichi Takeda for his many useful advices on the analysis of
our Subaru/HDS data, and for his opening the TGVIT and SPTOOL programs into public. 
When we conducted our solar observation at Hida Observatory mentioned in Section \ref{subsec:Hida}, 
we received important advices and kind supports from 
Dr. Satoru Ueno, Dr. Tetsu Anan, Dr. Reizaburo Kitai, Dr. Ayumi Asai, and other members of Kwasan and Hida Observatories, Kyoto University.
Kepler was selected as the tenth Discovery mission. 
Funding for this mission is provided by the NASA Science Mission Directorate. 
The Kepler data presented in this paper were obtained from the
Multimission Archive at STScI. 
This work was supported by the Grant-in-Aids from the Ministry of Education, 
Culture, Sports, Science and Technology of Japan (No. 25287039, 26400231, and 26800096).

\clearpage

\clearpage

  \begin{center}
\begin{longtable}{lccccccccc}
  \caption{Rotational velocity and activity index of Ca II IRT and H$\alpha$.}\label{tab:r0CaHa}
\hline
Starname& $v \sin i $ & $v_{\rm{lc}}$ & $P_{0}$\footnotemark[a] & $\langle$BVAmp$\rangle$\footnotemark[b] & $r_{0}(8542)$ & $\langle fB\rangle$\footnotemark[c] 
& $ r_{0}(8498) $  & $ r_{0}(8662) $ & $ r_{0}(\rm{H}\alpha) $ \\ 
 & [km s$^{-1}$] & [km s$^{-1}$] & [day] & [\%] & & [Gauss] & & &  \\
\hline
\endhead
\hline
\endfoot
\hline
\multicolumn{10}{l}{\hbox to 0pt{\parbox{165mm}{\footnotesize
    \footnotemark[a] Period values estimated from the Kepler data (Quarter 2$\sim$16 data), which are reported in Paper I.
}}}
\\
\multicolumn{10}{l}{\hbox to 0pt{\parbox{165mm}{\footnotesize
    \footnotemark[b] The amplitude of the brightness variation. This value is calculated by taking the average of the amplitude value of each Quarter (Q2$\sim$Q16) data.
(The amplitude value of each Quarter data is available in Supplementary Data.)
The errors of $\langle$BVAmp$\rangle$ correspond to the maximum and minimum of the amplitude value of all Quarter (Q2$\sim$Q16) data.
}}}
\\
\multicolumn{10}{l}{\hbox to 0pt{\parbox{165mm}{\footnotesize
    \footnotemark[c] Mean intensity of the stellar magnetic field estimated from $r_{0}(8542)$ index (See Section \ref{subsec:Hida} for details.).
}}}
\\   
\multicolumn{9}{l}{\hbox to 0pt{\parbox{165mm}{\footnotesize
    \footnotemark[d] We already reported the values of KIC6934317 in \citet{SNotsu2013}. }}}
\\ 
\multicolumn{9}{l}{\hbox to 0pt{\parbox{165mm}{\footnotesize
\footnotemark[e] We do not use Ca II IRT (8498/8542/8662) data of these stars 
(KIC7093547, KIC8359398, KIC8547383, KIC9459362, KIC10252382, KIC10387363, and KIC11818740) since S/N ratio of the observed spectra is too low.}}}
\\    
\multicolumn{9}{l}{\hbox to 0pt{\parbox{165mm}{\footnotesize
    \footnotemark[f] We already reported the $v \sin i$ and $r_{0}(8542)$ values of KIC9766237 and KIC9944137 in \citet{Nogami2014}. }}}    
\endlastfoot
KIC3626094 & 2.9 & 100.6 & 0.7 & $ 0.24^{+0.03}_{-0.05} $ & 0.24 & 0.6$\pm$6 & 0.33 & 0.25 & 0.21    \\ 
KIC4742436 & 3.6 & 28.3 & 2.3 & $ 0.46^{+0.43}_{-0.24} $ & 0.30 & 3$\pm$6 & 0.40 & 0.32 & 0.29    \\ 
KIC4831454 & 2.1 & 8.6 & 5.2 & $ 0.91^{+1.27}_{-0.50} $ & 0.50 & 68$\pm$41 & 0.58 & 0.44 & 0.32    \\ 
KIC6503434 & 5.5 & 39.5 & 3.9 & $ 0.22^{+0.18}_{-0.09} $ & 0.27 & 1$\pm$6 & 0.35 & 0.27 & 0.27    \\ 
KIC6504503 & 1.6 & 1.4 & 31.8 & $ 0.14^{+0.34}_{-0.11} $ & 0.25 & 0.8$\pm$6 & 0.35 & 0.26 & 0.21    \\ 
KIC6865484 & 2.7 & 4.5 & 10.3 & $ 0.72^{+0.40}_{-0.15} $ & 0.48 & 52$\pm$41 & 0.56 & 0.42 & 0.40    \\ 
KIC6934317 & 1.9 \footnotemark[d] & 19.7 & 2.5 & $ 0.14^{+0.03}_{-0.04} $ & 0.33 \footnotemark[d] & 5$\pm$24 & 0.43 \footnotemark[d] & 0.37 \footnotemark[d] & 0.42 \footnotemark[d]   \\ 
KIC7093547 \footnotemark[e] & 2.1 & 3.2 & 14.2 & $ 0.37^{+0.32}_{-0.19} $ & - & - & - & - & 0.32    \\ 
KIC7354508 & 3.2 & 4.5 & 16.8 & $ 0.50^{+0.32}_{-0.28} $ & 0.40 & 16$\pm$24 & 0.42 & 0.33 & 0.24    \\ 
KIC7420545 & 6.3 & 4.8 & 36.2 & $ 1.11^{+0.89}_{-0.56} $ & 0.40 & 16$\pm$24 & 0.50 & 0.37 & 0.32    \\ 
KIC8359398 \footnotemark[e] & 1.8 & 2.9 & 12.7 & $ 1.68^{+0.56}_{-0.96} $ & - & - & - & - & 0.34    \\ 
KIC8429280 & 37.1 & 31.6 & 1.2 & $ 3.71^{+1.21}_{-1.34} $ & 0.68 & 489$\pm$173 & 0.80 & 0.72 & 0.93    \\ 
KIC8547383 \footnotemark[e] & 4.4 & 4.0 & 14.8 & $ 0.66^{+0.38}_{-0.31} $ & - & - & - & - & 0.38    \\ 
KIC8802340 & 4.3 & 4.4 & 10.3 & $ 1.69^{+0.88}_{-0.83} $ & 0.49 & 60$\pm$41 & 0.57 & 0.46 & 0.31    \\ 
KIC9412514 & 8.1 & 38.9 & 3.7 & $ 0.04^{+0.01}_{-0.01} $ & 0.22 & 0.4$\pm$6 & 0.32 & 0.22 & 0.19    \\ 
KIC9459362 \footnotemark[e] & 7.0 & 9.3 & 12.6 & $ 1.78^{+1.22}_{-0.88} $ & - & - & - & - & 0.46    \\ 
KIC9583493 & 7.2 & 9.1 & 5.5 & $ 1.35^{+0.61}_{-0.47} $ & 0.51 & 77$\pm$123 & 0.61 & 0.49 & 0.38    \\ 
KIC9652680 & 34.2 & 29.0 & 1.5 & $ 4.72^{+1.32}_{-2.06} $ & 0.68 & 489$\pm$173 & 0.76 & 0.69 & 0.60    \\ 
KIC9766237 & 2.1 \footnotemark[f] & 4.2 & 14.2 & $ 0.09^{+0.04}_{-0.03} $ & 0.26 \footnotemark[f] & 1$\pm$6 & 0.40 & 0.28 & 0.23    \\ 
KIC9944137 & 1.9 \footnotemark[f] & 3.7 & 12.6 & $ 0.07^{+0.03}_{-0.02} $ & 0.27 \footnotemark[f] & 1$\pm$6 & 0.41 & 0.29 & 0.22    \\ 
KIC10252382 \footnotemark[e] & 5.5 & 6.1 & 16.8 & $ 2.93^{+3.95}_{-1.63} $ & - & - & - & - & 0.52    \\ 
KIC10387363 \footnotemark[e] & 1.5 & 1.4 & 29.9 & $ 0.30^{+0.35}_{-0.18} $ & - & - & - & - & 0.31    \\ 
KIC10471412 & 2.7 & 3.2 & 15.2 & $ 0.24^{+0.20}_{-0.12} $ & 0.35 & 7$\pm$24 & 0.48 & 0.34 & 0.28    \\ 
KIC10528093 & 11.4 & 8.5 & 12.2 & $ 1.94^{+1.19}_{-0.90} $ & 0.67 & 445$\pm$173 & 0.75 & 0.66 & 0.51    \\ 
KIC11140181 & 3.6 & 3.7 & 11.5 & $ 1.52^{+1.15}_{-0.62} $ & 0.56 & 141$\pm$123 & 0.63 & 0.45 & 0.36    \\ 
KIC11197517 & 0.9 & 3.2 & 14.3 & $ 0.10^{+0.18}_{-0.06} $ & 0.19 & 0.1$\pm$6 & 0.27 & 0.21 & 0.22    \\ 
KIC11303472 & 2.1 & 2.8 & 13.5 & $ 1.50^{+0.69}_{-0.75} $ & 0.49 & 60$\pm$41 & 0.56 & 0.39 & 0.33    \\ 
KIC11390058 & 3.5 & 4.2 & 12.0 & $ 0.33^{+0.12}_{-0.21} $ & 0.32 & 4$\pm$24 & 0.42 & 0.30 & 0.29    \\ 
KIC11455711 & 7.9 & 7.6 & 13.9 & $ 0.92^{+0.44}_{-0.42} $ & 0.40 & 16$\pm$24 & 0.59 & 0.44 & 0.30    \\ 
KIC11494048 & 3.4 & 3.4 & 14.8 & $ 0.30^{+0.13}_{-0.16} $ & 0.32 & 4$\pm$24 & 0.45 & 0.31 & 0.23    \\ 
KIC11610797 & 22.5 & 38.4 & 1.6 & $ 2.44^{+0.66}_{-0.95} $ & 0.63 & 300$\pm$173 & 0.70 & 0.61 & 0.46    \\ 
KIC11764567 & 14.7 & 3.6 & 22.4 & $ 1.74^{+0.85}_{-0.78} $ & 0.49 & 60$\pm$41 & 0.57 & 0.44 & 0.39    \\ 
KIC11818740 \footnotemark[e] & 3.0 & 2.7 & 15.4 & $ 1.40^{+0.81}_{-0.93} $ & - & - & - & - & 0.35    \\ 
KIC12266582 & 4.8 & 6.1 & 6.9 & $ 1.58^{+0.50}_{-0.55} $ & 0.52 & 88$\pm$123 & 0.61 & 0.48 & 0.37    \\ 
59Vir & 6.1 & - & - & - & 0.40 & 16$\pm$24 & 0.50 & 0.38 & 0.27    \\ 
61Vir & $<$1.0 & - & - & - & 0.19 & 0.1$\pm$6 & 0.30 & 0.20 & 0.18    \\ 
18Sco & 2.0 & - & - & - & 0.20 & 0.2$\pm$6 & 0.30 & 0.21 & 0.18    \\ 
HD163441 & 2.5 & - & - & - & 0.23 & 0.5$\pm$6 & 0.34 & 0.25 & 0.19    \\ 
HD173071 & 2.7 & - & - & - & 0.25 & 0.8$\pm$6 & 0.34 & 0.25 & 0.21    \\ 
HIP100963 & 2.4 & - & - & - & 0.22 & 0.4$\pm$6 & 0.34 & 0.22 & 0.20    \\ 
HIP71813 & 2.3 & - & - & - & 0.20 & 0.2$\pm$6 & 0.30 & 0.20 & 0.19    \\ 
HIP76114 & 1.8 & - & - & - & 0.21 & 0.3$\pm$6 & 0.32 & 0.20 & 0.19    \\ 
HIP77718 & 2.5 & - & - & - & 0.23 & 0.5$\pm$6 & 0.34 & 0.23 & 0.20    \\ 
HIP78399 & 2.5 & - & - & - & 0.21 & 0.3$\pm$6 & 0.32 & 0.21 & 0.18    \\ 
Moon & 2.4 & - & - & - & 0.20 & 0.2$\pm$6 & 0.33 & 0.22 & 0.19    \\ 
\hline
    \end{longtable}
  \end{center}

\clearpage

\appendix

\section{Discussion of chromospheric activities using ``Flux Method''.}\label{sec:excessFlux}
\begin{figure}[htbp]
 \begin{center}
  \FigureFile(120mm,120mm){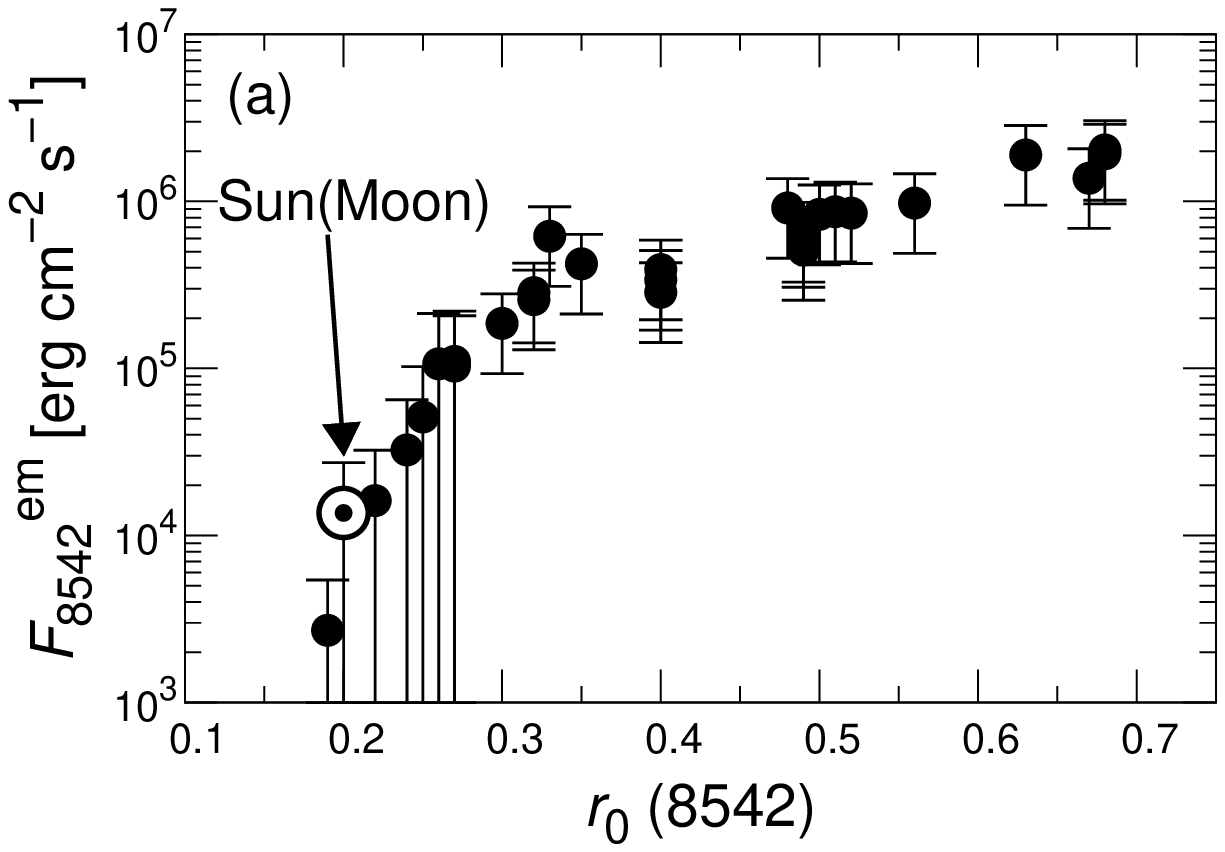}
  \FigureFile(120mm,120mm){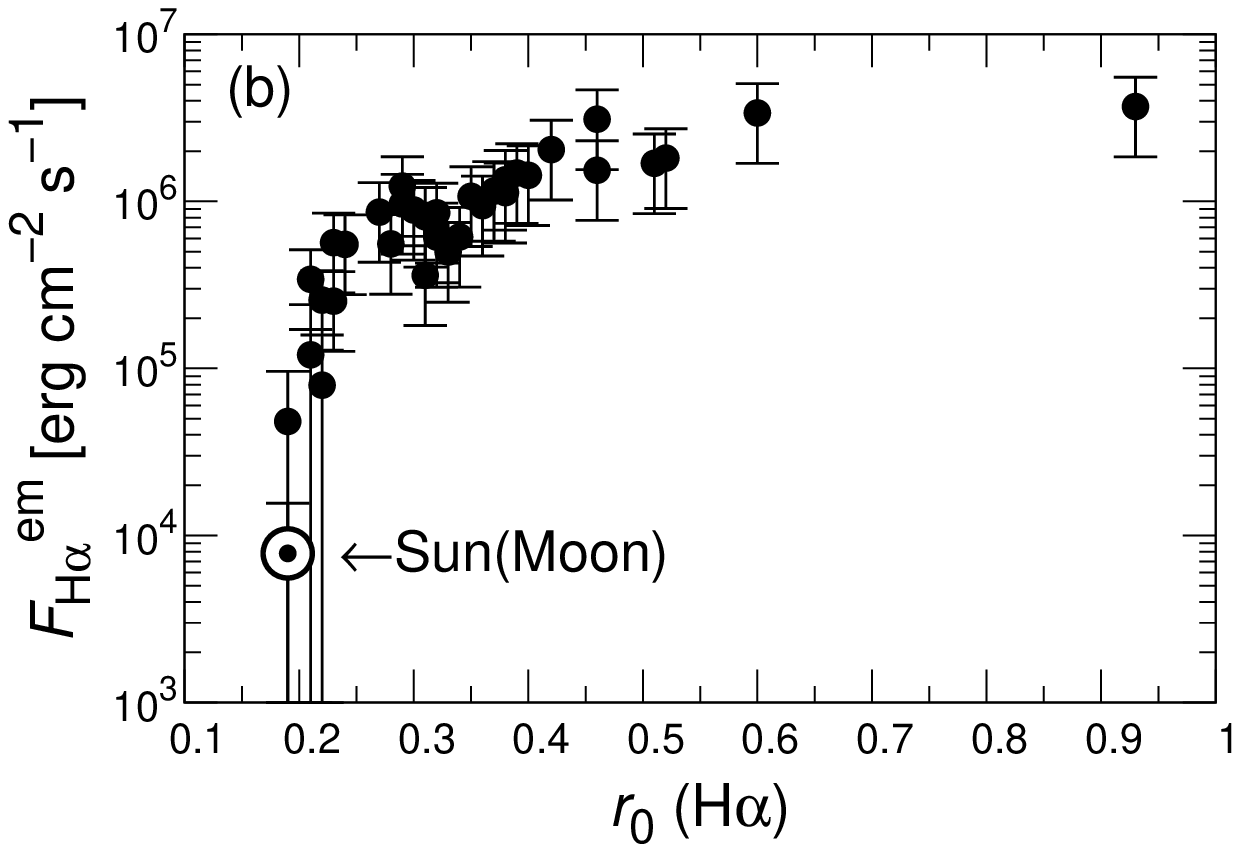}
 \end{center}
\caption{(a) $F^{\rm{em}}_{8542}$ as a function of $r_{0}$(8542). \\
(b)$F^{\rm{em}}_{H\alpha}$ as a function of $r_{0}$(H$\alpha$). \\
The Moon value in this paper is also plotted as a solar value for reference 
with a circled dot in both (a) and (b).
}\label{fig:flux}
\end{figure}

We measured the values of the $r_{0}$ of Ca II IRT and H$\alpha$ in Section \ref{subsec:CaIIHa}, 
and discussed chromospheric activity of superflare stars by using the $r_{0}$(8542) index in Section \ref{subsec:activity-spots}. 
This index is an indicator of chromspheric activity as described in the above sections, 
but it also depends on $v \sin i$ (e.g., Figure 5(a) of \cite{Takeda2010}).
A large value of $v \sin i$ can indeed increase the residual flux, mimicking
the effect of filling the line core with chromospheric emission.
Because of this, we here roughly estimated the emission flux of the Ca II 8542 and H$\alpha$ lines in order to remove this influence of $v \sin i$.
The ways of estimation is basically the same as done in \citet{SNotsu2013}, 
and we summarize the method in the following. 
\\ \\
\ \ \ \ \ \ \
We used the spectral subtraction technique (e.g., \cite{Frasca1994}; \cite{Frasca2011}; \cite{Martinez-Arnaiz2011}). 
With this subtraction process, we can subtract the underlying photospheric contribution from the spectrum of the star, and 
can investigate the spectral emission originating from the chromosphere in detail \citep{Martinez-Arnaiz2011}.
We used the spectrum of 61Vir obtained in this observation as an inactive template to be subtracted from the spectrum of the target stars. 
This is because 61Vir is a slowly rotating and non-active early G-type main sequence star (e.g., \cite{Anderson2010}).
We measured the excess equivalent width ($W^{\rm{em}}_{\lambda}$) of the Ca II 8542 and H$\alpha$ lines 
by using the residual spectrum around the line core resulting from this subtraction process.
We then derived emission fluxes ($F^{\rm{em}}_{\lambda}$) of Ca II 8542 and H$\alpha$ lines 
from $W^{\rm{em}}_{\lambda}$ of these lines by the following relation,
$F^{\rm{em}}_{\lambda}=W^{\rm{em}}_{\lambda}F^{\rm{cont}}_{\lambda}$ (\citet{Martinez-Arnaiz2011}),
where $F^{\rm{cont}}_{\lambda}$ is the continuum flux around the wavelength of each line. 
We calculated $F^{\rm{cont}}_{\lambda}$ by using the the following empirical 
relationships between $F^{\rm{cont}}_{\lambda}$ and color index ($B-V$) derived by \citet{Hall1996}:
 \begin{equation}\label{eq:FcontHa}
 \log F^{\mathrm{cont}}_{\mathrm{H}\alpha} = [7.538 - 1.081(B-V)]\pm 0.033, \ +0.0\leq B-V \leq +1.4,
 \end{equation}
 \begin{equation}\label{eq:FcontCaIRT1}
 \log F^{\mathrm{cont}}_{\mathrm{IRT}} = [7.223 - 1.330(B-V)]\pm 0.043, \ -0.1\leq B-V \leq +0.22,
 \end{equation}
 \begin{equation}\label{eq:FcontCaIRT2}
 \log F^{\mathrm{cont}}_{\mathrm{IRT}} = [7.083 - 0.685(B-V)]\pm 0.055, \ +0.22\leq B-V \leq +1.4.
 \end{equation}
For deriving $B-V$ values, we used the following empirical relations of low-mass main-sequence stars (F0V-K5V) 
among $B-V$, $T_{\rm{eff}}$, and [Fe/H] derived by \citet{Alonso1996}: 
\begin{eqnarray}\label{eq:TeffB-V}
(B-V)=-0.9837+1.8652\theta_{\rm{eff}}-0.0324\theta_{\rm{eff}}^{2}+0.1293\theta_{\rm{eff}}\rm{[Fe/H]} \nonumber \\
-0.0085\rm{[Fe/H]}+0.0199\rm{[Fe/H]}^{2}, \ \ \ \sigma(B-V)=0.039 \rm{mag}~,
\end{eqnarray}
where $\theta_{\rm{eff}}=5040/T_{\rm{eff}}$.
\\ \\ 
\ \ \ \ \ \ \ 
The estimated values of the emission flux of Ca II 8542 ($ F^{\rm{em}}_{8542}$) and H$\alpha$ ($F^{\rm{em}}_{\rm{H}\alpha}$) are listed 
in Supplementary Table 1 of this paper.
The excess equivalent width values ($W^{\rm{em}}_{8542}$ and $W^{\rm{em}}_{\rm{H}\alpha}$) are also listed in this table.
As we have already mentioned in Section \ref{subsec:CaIIHa}, 
we did not use the values of Ca II 8542 of 7 target superflare stars 
(KIC7093547, KIC8359398, KIC8547383, KIC9459362, KIC10252382, KIC10387363, and KIC11818740), which are marked 
in the 8th column of Supplementary Table 1.
This is because the exposure time for such stars was not sufficient and the resultant S/N ratio around Ca II triplet lines is low.
\\ \\
\ \ \ \ \ \ \
In Figure \ref{fig:flux}, the $F^{\rm{em}}_{8542}$ and $F^{\rm{em}}_{\rm{H}\alpha}$ are plotted as a function of $r_{0}$(8542) and $r_{0}$(H$\alpha$), respectively.
The error values of $F^{\rm{em}}_{8542}$ and $F^{\rm{em}}_{\rm{H}\alpha}$ in Figure \ref{fig:flux} are 
plotted in the following method. 
Roughly assuming the upper limit of subtraction error, 
we here consider the typical error value of the emission flux ($F^{\rm{em}}$) is roughly $\pm50$\%~, 
if excess equivalent width ($W^{\rm{em}}$) is larger than 30m\AA~(sufficiently high compared to 61Vir).
On the other hand, if the excess equivalent width ($W^{\rm{em}}$) is less than 30m\AA~(not so high compared to 61Vir), 
we consider that the typical error value of the emission flux ($F^{\rm{em}}$) is about $\pm100$\%~.
We can see a positive correlation between emission flux and $r_{0}$ index in this figure,  
and all of the superflare stars with high $r_{0}$(8542) and $r_{0}$(H$\alpha$) values show high $F^{\rm{em}}_{8542}$ and $F^{\rm{em}}_{\rm{H}\alpha}$ values.
This positive correlation show us that by using values of the emission flux,
we can conclude the basically same conclusions as we did with $r_{0}$(8542) index in Section \ref{subsec:activity-spots}.

\end{document}